\documentclass[showpacs,prd,reprint,preprintnumbers,amsmath,amssymb,nofootinbib]{revtex4-1}

\usepackage{graphicx,subfigure,multirow}
 \usepackage{longtable}
\usepackage{dcolumn}
\usepackage{bm}
\usepackage[all]{xy}
\usepackage{color}
\usepackage{slashed}
\usepackage[usenames,dvipsnames]{xcolor}
\usepackage[unicode=true,pdfusetitle,
 bookmarks=true,bookmarksnumbered=false,bookmarksopen=false,citecolor=Turquoise,
 breaklinks=false,pdfborder={0 0 1},backref=false,colorlinks=true]
 {hyperref}
 
 \usepackage{booktabs}
\usepackage{mathrsfs}
\usepackage{epsfig}
\usepackage{graphicx}
\usepackage{subfigure}
\usepackage{dcolumn}
\usepackage{bm}
\usepackage{amsmath}
\usepackage{slashed}
\usepackage{multirow}
\usepackage{color}
\usepackage{dcolumn}
\usepackage{bm}
\usepackage{color}

 \newcommand{\lsim}{{\;\raise0.3ex\hbox{$<$\kern-0.75em\raise-1.1ex\hbox{$\sim$}}\;}}
\newcommand{\gsim}{{\;\raise0.3ex\hbox{$>$\kern-0.75em\raise-1.1ex\hbox{$\sim$}}\;}}
\newcommand{\beq}{\begin{equation}}
\newcommand{\eeq}{\end{equation}}
\newcommand{\bea}{\begin{eqnarray}}
\newcommand{\eea}{\end{eqnarray}}

\def\baa{\begin{array}}
\def\eaa{\end{array}}

\mathchardef\minus="002D

\newcommand{\tabincell}[2]{\begin{tabular}{@{}#1@{}}#2\end{tabular}}

\begin{document}

\title{Hunting for top partner with a new signature at the LHC}

\author{Daohan Wang$^{1,2,3}$}
\email{First author}
\author{Lei Wu$^{4}$}
\email{Co-corresponding author: leiwu@njnu.edu.cn}
\author{Mengchao Zhang$^{1}$}
\email{Corresponding author: mczhang@jnu.edu.cn}

\vspace{0.5cm}

\affiliation{$^1$Department of Physics and Siyuan Laboratory, Jinan University, Guangzhou 510632, P.R. China}
\affiliation{$^2$CAS Key Laboratory of Theoretical Physics, Institute of Theoretical
                Physics, Chinese Academy of Sciences, Beijing 100190, P. R. China}
\affiliation{$^3$School of Physical Sciences, University of Chinese Academy of Sciences,Beijing 100049, P. R. China}
\affiliation{$^4$Department of Physics and Institute of Theoretical Physics, Nanjing Normal University, Nanjing 210023, P. R. China}

\vspace{0.5cm}

\date{\today}

\begin{abstract}
Vector-like top partner plays a central role in many new physics models which attempt to address the hierarchy problem. The top partner is conventionally assumed to decay to a quark  and a SM boson. However, it is also possible that the top partner decays to a non-SM scalar and a top quark. Such an exotic decay channel can be the main decay channel of top partner, and thus provides new windows to search for top partner at the LHC. In this paper, with classical machine learning method Boosted Decision Tree (BDT), we perform a model independent study of the discovery potential of this new signature at the LHC. In order to suppress the main QCD background, we consider subdominant but clean decay channel $a\to \gamma\gamma$. For completeness, the single production process and pair production process of top partner are all taken into account. We find that, for both single and pair production, the future High-Luminosity LHC can exclude the top partner mass up to TeV scale through channel $T\to t a$ ($a\to \gamma\gamma$), even if $BR(a\to \gamma\gamma)$ is as small as $\mathcal{O}$(0.1\%). Besides, our result shows that single production can overmatch pair production at 14 TeV LHC, provided that top partner is heavier than $800\sim 900$ GeV. 

\end{abstract}



\maketitle

\section{Introduction}
\label{sec:introduction}

The observation of Higgs boson has completed the Standard Model (SM)~\cite{Aad:2012tfa,Chatrchyan:2012xdj}. However, the radiative stability of the Higgs boson mass is widely considered as a major theoretical motivation for new physics beyond the SM.
A popular method to cure this problem is to introduce a softly-broken supersymmetry (SUSY), then quadratically divergences can be cancelled exactly by the super-partners. Due to large top Yukawa coupling, spin-0 top-squark plays a central role in SUSY search. 
Besides, spin-1/2 vector-like top partner (noted as $T$) can also arise in new physics models which attempt to stabilize the Higgs mass, like the composite Higgs model with partial compositeness~\cite{Kaplan:1983fs,Kaplan:1991dc,Redi:2012ha,Contino:2006qr,Matsedonskyi:2012ym,Marzocca:2012zn,Blasi:2019jqc,Blasi:2020ktl}.
Through the mixing with top quark, spin-1/2 vector-like top partner decays to $bW^+$, $tZ$ and $th$~\cite{Agashe:2004rs,simone2012partner}. 
Current direct searches, which are designed for these conventional decay channels, have excluded the mass of top partner up to about 1 TeV~\cite{Aaboud:2018xuw, Aaboud:2018saj, Aaboud:2017qpr, Aaboud:2017zfn, Aaboud:2018wxv, Aaboud:2018uek, Aaboud:2018pii, Sirunyan:2018omb, Sirunyan:2017pks, Sirunyan:2019sza,Han:2014qia,Liu:2015kmo}.


\begin{figure}[ht]
\centering
\includegraphics[width=8.5cm]{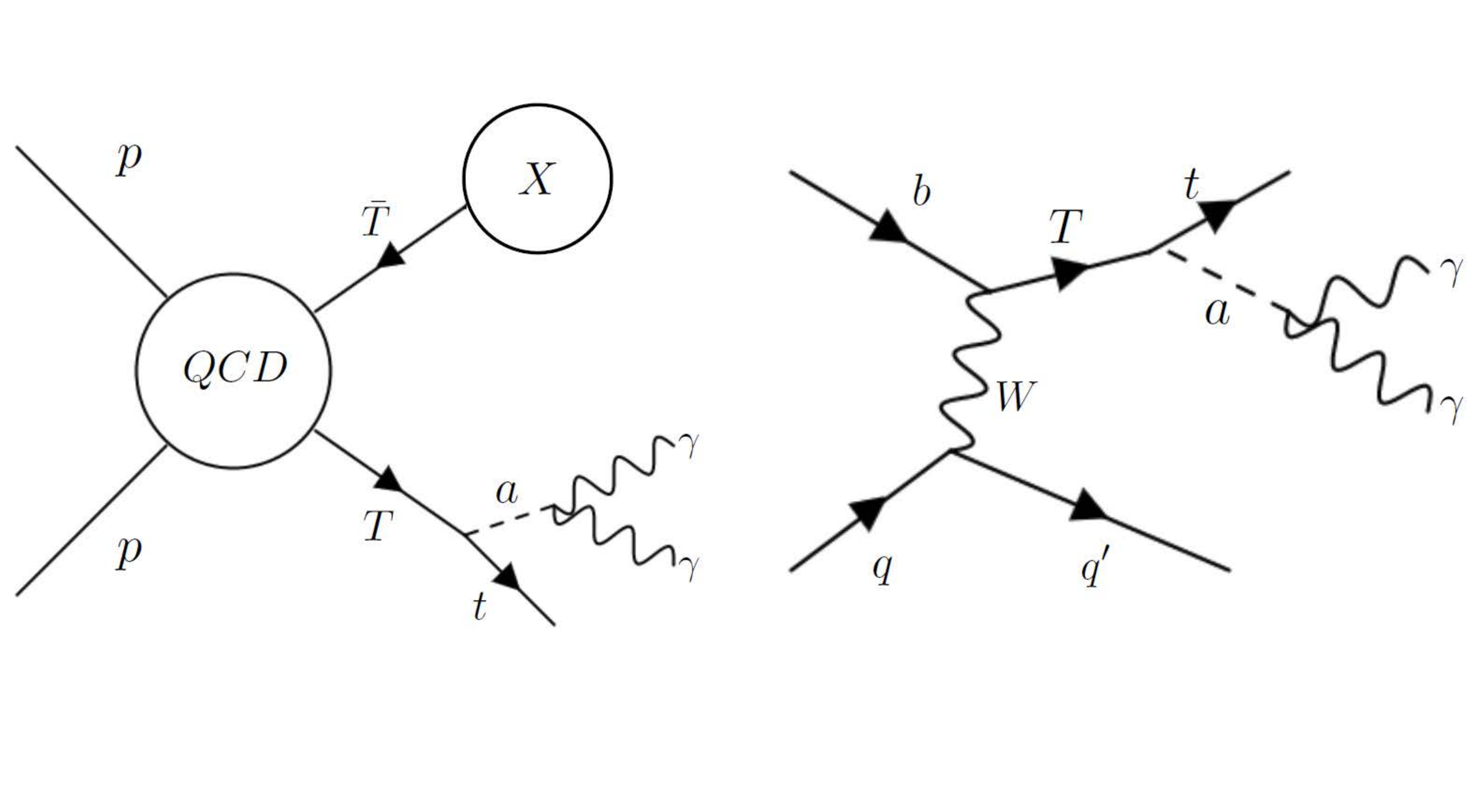} 
\vspace{-1cm}
\caption{Feynman Diagrams: the pair production process $pp \to T\bar{T}$ ({\it left}), and the single production process $pp \to Tj$ ({\it right}) at the LHC. 
Bottom quark is considered as a parton in proton.}
\label{production}
\end{figure}

However, it is possible for vector-like top partner to decay exotically~\cite{Anandakrishnan:2015yfa,Kraml:2016eti,Bizot:2018tds,Serra:2015xfa,Han:2018hcu,Alhazmi:2018whk,Benbrik:2019zdp,Aguilar-Saavedra:2017giu,Chala:2017xgc,Cacciapaglia:2019zmj,Kim:2018mks,Xie:2019gya,Criado:2019mvu,Aguilar-Saavedra:2019ghg,Ramos:2019qqa}. 
For example, if the ultraviolet (UV) completion of composite Higgs model is constructed by introducing new fermions that are charged under a new strong gauge interaction, then generally there are other light pNGBs in addition to the SM Higgs doublet~\cite{Barnard:2013zea,Ferretti:2013kya,Ferretti:2016upr,DeGrand:2016pgq}. 
In such UV construction, a light CP-odd pseudo-scalar (noted as $a$), which is associated to a non-anomalous axial $U(1)$ global symmetry, always arise.
This can lead to a new decay channel of vector-like top quark, $T \to t a$, if this is allowed in kinematics~\footnote{$a$ or other pNGBs can also be directly probed by searching for di-boson or fermion pair signals~\cite{Belyaev:2016ftv,Cacciapaglia:2019bqz}.}. If $a$ is heavier than 350 GeV, it mainly decays to $t\bar{t}$ and results  in six top quarks final states~\cite{Han:2018hcu}. If $a$ is lighter than 350 GeV, its dominant decay channel can be $a \to b\bar{b}$ or $a \to gg$. In the former case, due to multiple b-jets in final state, current data can exclude $m_T$ up to about 1 TeV~\cite{Cacciapaglia:2019zmj}. While in the latter case, the huge QCD background greatly reduce the sensitivity of current LHC searches, and $m_T$ around 400 GeV $\sim$ 550 GeV can still survive under current direct search bounds~\cite{Cacciapaglia:2019zmj}.

To overcome the difficulties in the case of $BR(a\to gg) \approx 1$, in this paper we consider the subdominant but much cleaner decay channel of pNGB $a$, $a\to \gamma\gamma$. 
Different from previous work~\cite{Benbrik:2019zdp}, we will adopt the classical machine learning method Boosted Decision Tree (BDT)~\cite{Roe:2004na} to improve the search sensitivity\footnote{Other machine learning methods have been used in top partner search, e.g.~\cite{Romao:2019dvs}.}, and focus on the single production process of $T$, $pp \to Tj$ (c.f. Fig.~\ref{production}({\it right})). 
The pair production process of $T$, $pp\to \bar{T}T$, will also be considered as an comparison of single production process (c.f. Fig.~\ref{production}({\it left}))\footnote{In composite Higgs model, top partner could have more exotic production process, e.g.~\cite{Araque:2015cna,Dasgupta:2019yjm}.}. 
Due to the large QCD coupling, $pp\to \bar{T}T$ is the conventional production process in top partner search.
In contrast with the pair production process, $pp \to Tj$ is induced by electro-weak coupling which is much weaker than QCD coupling, and thus its crosssection is generally considered to be negligible.
But single production of top partner has a larger phase space and can be enhanced by the collinear effect from the light quark emitting a $W$-boson in high energy region~\cite{Willenbrock:1986cr}. These features may make the single production process as a sensitive probe of top partner at the LHC, especially when top partner is heavy\footnote{Recent study of singly produced vector-like quarks see~\cite{Roy:2020fqf}.}.
In addition, Our analysis will be performed in a model independent way, and can be easily interpreted to a concrete model. 

The rest of this paper is organized as follows. In Section II, we present the model framework, which is the SM extended by a vector-like top partner $T$ and a light pNGB $a$. 
In Section III we perform a detailed Monte Carlo simulation of our signal process and main background process. 
Model independent exclusion limits for single and pair production process, and the search sensitivity comparison, will be given. 
In Section IV we use two benchmark models to show how to obtain the exclusion limits for concrete models by using our model independent results.  
Finally we conclude our work in Section V.

\section{Model Framework}
\label{sec2}
We consider the SM extended by a vector-like top partner $T$ (with electric charge $+2/3$) and a light pseudo-scalar $a$.
This simplified scenario can be embedded in many new physics models~\cite{Contino:2004vy,Cacciapaglia:2008bi,Schmaltz:2002wx,Aguilar-Saavedra:2017giu,Benbrik:2019zdp,Chala:2017xgc,Dermisek:2019vkc,Ferretti:2016upr}. To be specific, the relevant Lagrangian of the vector-like top partner can be expressed as
\begin{eqnarray}
\mathcal{L}_{T} &=& \phantom{+}  \overline{T}\left(i\slashed{D}-m_T\right) T + \left(\kappa^T_{W,L} \frac{g}{\sqrt{2}}  \,\overline{T}\slashed{W}^+P_Lb     \right.    \nonumber \\ 
&&\left.  + \kappa^T_{Z,L} \frac{g}{2 c_W}\, \overline{T} \slashed{Z} P_L t - \kappa^T_{h,L} \frac{m_T}{v}\, \overline{T} h P_L t \right.    \nonumber \\ 
&&\left.  + i \kappa^T_{a,L} \frac{m_T}{v} \, \overline{T} a P_L t  + L\leftrightarrow R + \mbox{ h.c. }\right), 
\label{eq:LTa}
\end{eqnarray}
where $m_T$ is the mass of top partner, and couplings $\kappa^T_i$ describe the effective interactions between the top partner and other particles. 
Since we are interested in the single top partner production process $pp \to T (\to ta) j $, which comes from the electroweak coupling, we only keep $\kappa^T_{W,L}$ and $\kappa^T_{a,L}$\footnote{For conciseness, $\kappa^T_{W,L}$ and $\kappa^T_{a,L}$ will be noted as $\kappa^T_{W}$ and $\kappa^T_{a}$ in the rest of this paper.}, and neglect other $\kappa^T_i$ in this work for simplicity. 
The appearance of $\kappa^T_{W}$ will inevitably lead to conventional decay channel $T\to bW^+$. 
Current direct search~\cite{Aaboud:2017zfn} already exclude the mass of $T$ up to 1.1 TeV, provided $BR(T\to bW^+)$ is 0.5. 
However, $\kappa^T_{W}$ is determined by $SU(2)$ gauge coupling and the mixing angle between top partner and elementary top~\cite{Bizot:2018tds}. 
But $\kappa^T_{a}$ is induced by the coupling between top partner and pNGB $a$, which can come from a strong interaction sector.
So $\kappa^T_{a}$ can be much larger than $\kappa^T_{W}$, and makes $T\to ta$ the main decay channel.  
$BR(T\to bW^+)$ can be small enough to escape current direct search bounds. 
Concrete models and discussion will be given in Sec.~\ref{sec4}.

On the other hand, the Lagrangian of pseudo-scalar $a$ and its interactions with SM particles can be expressed as:
\begin{eqnarray}
	{\mathcal{L}}_a &=& \frac{1}{2}(\partial_\mu a)(\partial^\mu a) 	-\frac{1}{2} m_a^2 a^2 +  \sum_f \frac{i C_f^a m_f}{f_a} a \bar f \gamma^5 f \nonumber  \\
    & & + \frac{g_s^2 K_g^a }{16\pi^2 f_a}a {G}^a_{\mu\nu}\tilde{{G}}^{a\mu\nu} + \frac{e^2 K^a_\gamma}{16\pi^2 f_a} a A_{\mu\nu} \tilde{A}^{\mu\nu} \nonumber   \\
	& &  + \frac{g^2 c_W^2 K^a_Z}{16 \pi^2 f_a} a Z_{\mu\nu} \tilde{Z}^{\mu\nu} + \frac{egc_W K^a_{Z\gamma}}{8\pi^2 f_a} a A_{\mu\nu} \tilde{Z}^{\mu\nu} \nonumber  \\
      & &+ \frac{g^2 K^a_W}{16 \pi^2 f_a} a W_{\mu\nu} \tilde{W}^{\mu\nu}, 
	\label{eq:aLag}
\end{eqnarray}
where the gauge boson field strength and its conjecture are noted as $V_{\mu\nu} \equiv \partial_\mu V_\nu - \partial_\nu V_\mu$ and $\tilde{V}_{\mu\nu} \equiv \epsilon_{\mu\nu\rho\sigma}V^{\rho\sigma}$. 
${G}^a_{\mu\nu}$, $A_{\mu\nu}$, $Z_{\mu\nu}$, and $W_{\mu\nu}$ are the field strength of gluon, photon, $Z$ boson, and $W$ boson.
$m_a$ and $m_f$ are the mass of pseudo-scalar $a$ and SM fermions. $C^a_f$ and $K^a_V$ are dimensionless coupling coefficients of $a\bar{f}f$ and $aV\tilde{V}$ terms. $g_s$, $e$, $g$, and $c_W$ are strong coupling constant, electric charge, $SU(2)$ coupling constant, and the cosine of Weinberg angle respectively.

If $C^a_f$ are comparable with $K^a_V$, then $BR(a\to\bar{b}b)$ will be a dominant decay channel of $a$ when $m_a < 350$ GeV. 
In this case, multiple bottom quark jets in the final state help to suppress QCD background, and $m_T \lesssim 1$ TeV can be excluded by current data~\cite{Cacciapaglia:2019zmj}. 
However, if $C^a_f$ is negligible, $a\to gg$ will be the dominant decay channel and current direct search become insensitive, especially when $m_a \lesssim 100 $GeV~\cite{Cacciapaglia:2019zmj}. 
This is because it is difficult for us to identify the jet pair coming from $a$ decay in the huge QCD background. 
To enhance the search sensitivity, we can consider the minor decay channel $a\to \gamma\gamma$.
Due to the hierarchy between $g_s^2$ and $e^2$, $BR(a\to\gamma\gamma)$ is generally much smaller than $BR(a\to gg)$ in most models. 
But the photon pair signature from this decay channel is very clean, and it helps to suppress the QCD background greatly. 
In the rest of this paper we will assume $C^a_f$ to be negligible and focus on photon pair signature in the collider analysis.

\section{Collider Simulation and Analysis}
\label{sec3}
\begin{figure}[ht]
\centering
\includegraphics[width=9cm]{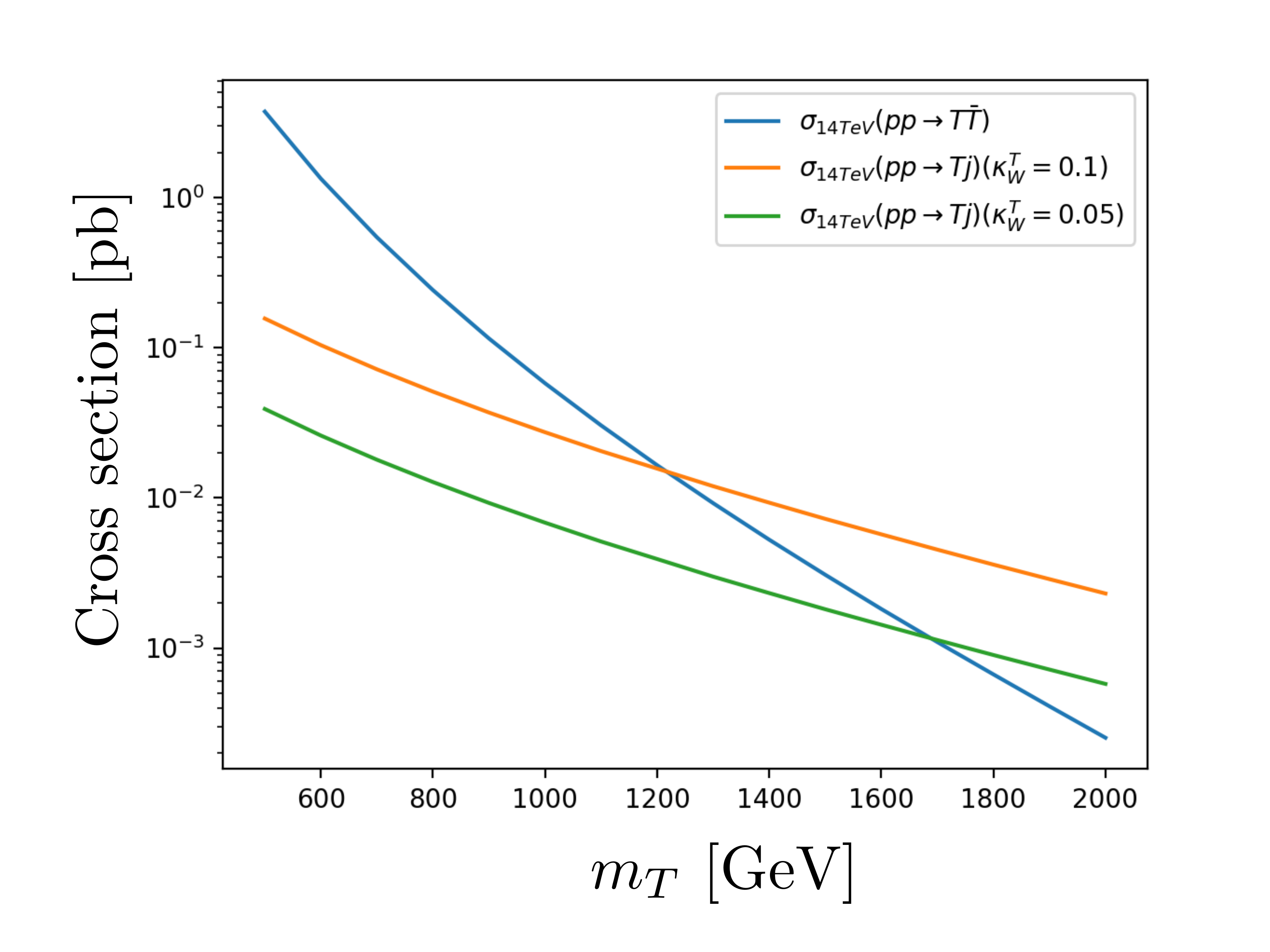} 
\vspace{-0.5cm}
\caption{The cross sections of single $T$ and  $T$ pair production at 14 TeV LHC. The conjugated process $pp \to \bar{T}j$ is included in single production as well. The coupling $k^T_W$ is set to 0.1 and 0.05 as examples.}
\label{cx}
\end{figure}

In Fig.~\ref{cx}, we present the cross sections of single and pair production of top partner at 14 TeV LHC. 
We normalize the leading order (LO) results, which are calculated by {\tt MadGrpaph5}~\cite{Alwall:2011uj}, to the next leading order (NLO) QCD predictions by using  $K$-factors 0.95~\cite{Cacciapaglia:2018qep} and 1.3~\cite{Fuks:2016ftf} for single and pair production process, respectively. 
It can be seen that, as $m_T$ increases, the cross section of $T$ pair production decreases faster than the cross section of single $T$ production. 
In the parameter region where $m_T$ is greater than 1.2 TeV (1.7 TeV),  the single production can have a cross section larger than the pair production at 14 TeV LHC, when the value of $k^T_W$ is 0.1 (0.05).

In our basic setting, we only maintain top partner's coupling $\kappa^T_{W}$ and $\kappa^T_{a}$, and thus $T$ can only decay to $ta$ or $bW^+$.
As we explained in Sec.~\ref{sec2}, we will only consider decay channel $T\to ta$ and neglect $T\to bW^+$. 
In order to suppress the QCD background, for both single and pair production process, we require a photon pair to appear in the final state. 
Thus for single production, the only process we need to consider is $pp\to Tj \to taj$ followed by $a\to \gamma\gamma$.
For pair production, due to a generally very small $BR(a\to \gamma\gamma)$, the signal process we consider is $pp\to T\bar{T}\to t\bar{t}aa$, followed by one $a$ decays to $\gamma\gamma$ and another $a$ decays to $gg$. 
After such choice, we can treat production cross-sections, $BR(T\to ta)$, and $BR(a\to\gamma\gamma)$ as undetermined parameters, and focus on the kinematic variables' distribution\footnote{$BR(a\to gg)$ is almost equal to 1 after we assuming $C^a_f$ to be negligible in Lagrangian~\ref{eq:aLag}.}. 
Thus for a model independent analysis, only $m_T$ and $m_a$ are relevant.


For Monte Carlo simulation, we implement our effective Lagrangian to an UFO model file~\cite{Degrande:2011ua} by {\tt FeynRules}~\cite{Alloul:2013bka}, and generate the parton-level signal and background events with {\tt MadGrpaph5}~\cite{Alwall:2011uj}. Parton shower and hadronization are performed by {\tt PYTHIA}~\cite{Sjostrand:2014zea}. The detector effect is simulated by {\tt Delphes}~\cite{deFavereau:2013fsa}. We assume the $b$-tagging efficiency to be $70\%$ and the rate of mis-tagging a light quark jet or gluon jet as a $b$-jet to be $1\%$. 
A jet might be mistagged as a photon in hadron collider environment. We use the jet faking photon rate given in~\cite{ATLAS:2016ukn} to estimate this effect.

\subsection{single production of top partner}

\begin{table*}[t!]
\begin{center}\begin{tabular}{|c|c|c|c|c|c|c|c|c|}
\hline  Basic Selection & \tabincell{c}{BP1 \\ (fb)} & \tabincell{c}{BP2 \\ (fb)} & \tabincell{c}{$t\bar{t}h$\\$(h\to\gamma\gamma)$ \\ (fb)} & \tabincell{c}{ $Whjj$\\$(h\to\gamma\gamma)$ \\ (fb)} & \tabincell{c}{ $t\bar{t}\gamma\gamma$ \\ (fb)}  &\tabincell{c}{ $tj\gamma\gamma$ \\ (fb)} & \tabincell{c}{ $Wjj\gamma\gamma$ \\ (fb)} & \tabincell{c}{ $thj$\\$(h\to\gamma\gamma)$ \\ (fb) }   \\
\hline \tabincell{c}{hardest jet with \\ $p_\text{T} > 100$ GeV \\ and $|\eta| < 2.5$} &0.521  &0.524 & 0.801 & 0.415 & 10.2 & 14.1 & 165.8 & 0.0749  \\
\hline \tabincell{c}{1 b-jet with \\ $p_\text{T} > 20$ GeV \\ and $|\eta| < 2.5$} &0.502 &0.502 & 0.372 & 0.0422 & 4.73 & 5.37 & 19.1 & 0.0300   \\
\hline \tabincell{c}{1 lepton with \\ $p_\text{T} > 20$ GeV \\ and $|\eta| < 2.5$}&0.108 &0.107 & 0.116 & 0.0076 & 1.72 &1.18 & 3.29 & 0.0030 \\
\hline  \tabincell{c}{2 $\gamma$ with \\ $p_\text{T} > 20$ GeV \\ and $|\eta| < 2.5$} &0.0382 &0.0336 & 0.0729 & 0.0054 &0.552 & 0.392 & 1.29 & 0.0021   \\
\hline \end{tabular} \caption{Cut-flow table of our basic selection criteria for single production process.  }
\label{Cutflow1}
\end{center}
\end{table*}

The full signal process we consider is $pp\to Tj\to t(\to bl^+\nu_l)a(\to\gamma\gamma)j$. 
The SM background are from the resonant processes: $pp\to t\bar{t}h$, $pp\to Whjj$, and $pp\to thj$, where the SM Higgs decays to photon pair. 
Besides, there are also non-resonant backgrounds: $pp\to t\bar{t}\gamma\gamma$, $pp\to tj\gamma\gamma$, and $pp\to Wjj\gamma\gamma$,
where two photons come from the radiation of charged particles. 
Each background process, e.g. $pp\to t\bar{t}\gamma\gamma$, can be faked by $pp\to t\bar{t}j\gamma$ ($pp\to t\bar{t}jj$) with one (two) hard jet mistagged as photon. We found that, due to a low jet faking photon rate (generally smaller than 0.05\%~\cite{ATLAS:2016ukn}), the cross section of process with faked photon is at least one order of magnitude smaller than the cross section of process without faked photon. 
So we neglect the jet faking photon effect in the rest of this paper, and only consider $pp\to t\bar{t}h$, $pp\to Whjj$, $pp\to thj$, $pp\to t\bar{t}\gamma\gamma$, $pp\to tj\gamma\gamma$, and $pp\to Wjj\gamma\gamma$ as our background processes\footnote{Pure QCD process like $pp\to t\bar{t}jj$ has a cross section which is much larger than that of $pp\to t\bar{t}\gamma\gamma$. So turning off the negligible QCD process like $pp\to t\bar{t}jj$ help us save lots of simulation time.}.

We use a basic selection criteria to select the events used in our analysis:
\begin{itemize}
\item[1.] The hardest jet is required to have $p_\text{T} > 100$ GeV and $|\eta| < 2.5$.
\item[2.] Exactly one b-jet with $p_\text{T} > 20$ GeV and $|\eta| < 2.5$.
\item[3.] Exactly one isolated lepton (electron or muon) with $p_\text{T} > 20$ GeV and $|\eta| < 2.5$.
\item[4.] Exactly two isolated photons with $p_\text{T} > 20$ GeV and $|\eta| < 2.5$.
\end{itemize}
The jet recoiled with top partner is generally quite hard, this is the reason for requiring the hardest jet in final state to have a large $p_\text{T}$. 
Other selection criteria just require the particles in the final state of signal process to exist. 
Here we need to emphasize that the cone size used in photon isolation is $\Delta R = 0.2$, and thus the signal event with two final state photons being too close to each other will be discarded.

To design a cut flow that can be used to enhance search significance, we need to study the kinematics of the final states for both signal and background process.  
To illustrate the distribution of kinematics variables, we consider two benchmark points (BP):
\begin{eqnarray}
\text{BP 1: } m_T &=& 800\text{ GeV} \ , \ m_a = 50\text{ GeV} \\ 
\text{BP 2: } m_T &=& 800\text{ GeV} \ , \ m_a = 200\text{ GeV} 
\end{eqnarray}
We also fix $\kappa^T_W$, $BR(T\to ta)$, and $BR(a\to\gamma\gamma)$  to 0.1, 100\%, and 1\% respectively\footnote{For a concrete model, $BR(T\to ta)$ surely can not be 100\%. In this section we only perform a model independent analysis and focus on the variables distribution.}. 
We use this setting as an example to show the crosssection of signal and background after the basic selection. 
The cut-flow table of our basic selection for these two BPs and the main backgrounds are given in Tab.~\ref{Cutflow1}. 

\begin{figure*}[t!]
\centering
\includegraphics[width=6.2in]{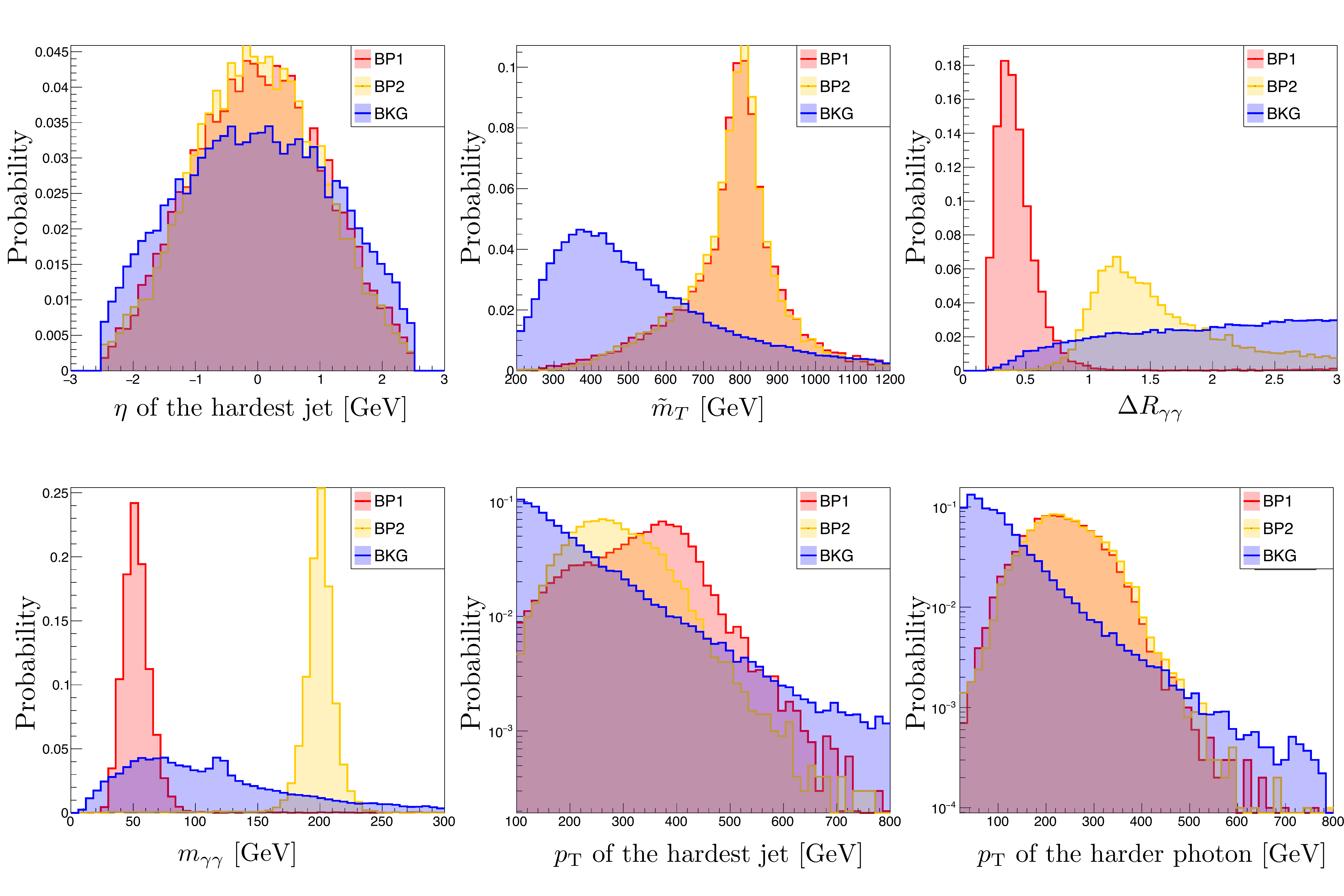}
\caption{Normalized distributions of the variables that are used to distinguish signal from background. Here BKG (background) means the combination of all the six background process. The proportion of each background process is proportional to their cross-section after the basic selection.}
\label{variables1}
\end{figure*}

After the basic selection, the type and number of final state particles are the same for signal and background events. 
While the cross section of the background is still about two orders of magnitude larger than the cross section of the signal process. 
To further suppress the background and enhance search sensitivity, we need to study the kinematics for both signal and background.
Here we study following variables that we think can be used in signal and background discrimination
\begin{itemize}
\item Pseudo-rapidity $\eta$ of the hardest jet. The hardest jet in signal process is evolved from the parton recoiled with the heavy top partner $T$, 
while the hardest jet in background process comes from parton radiation. 
Thus the hardest jet in signal process tends to be more central than the hardest jet in background process. 

\item We define a new variable which is called "reconstructed top partner mass" and noted as $\tilde{m}_T$:
\begin{eqnarray}
\tilde{m}_T = \sqrt{( p_{\text{visible}} + p_{\text{invisible}} )^2}
\end{eqnarray}
Here, the visible 4-momentum $p_{\text{visible}}$ is the sum of 4-momentums of b-jet, lepton, and 2 photons. 
Invisible 4-momentum $p_{\text{invisible}}$ is defined as $({E\!\!\!/}_{\text{T}} , \vec{p}^{\text{miss}}_{\text{T}} , 0)$.
The difference between $p_\nu$ and $p_{\text{invisible}}$ is the longitudinal momentum of neutrino. 
Due to a quite long decay chain of $T$, the momentum carried by neutrino is not so large compared with $m_T$.  
Thus missing the longitudinal momentum of neutrino will not change the reconstruction of top partner mass significantly, and we can expect $\tilde{m}_T \sim m_T$ for signal process.

\item Distance between two photons $\Delta R_{\gamma\gamma}$, which is defined as:
\begin{eqnarray}
\Delta R_{\gamma\gamma} = \sqrt{(\Delta\eta_{\gamma\gamma})^2 + (\Delta\phi_{\gamma\gamma})^2  }
\end{eqnarray}
$\Delta\eta_{\gamma\gamma}$ and $\Delta\phi_{\gamma\gamma}$ are the pseudorapidity and azimuthal angle difference of two photons. 
Because $a$ is highly boosted when $T$ is much heavier than $a$, so $\Delta R_{\gamma\gamma}$ tends to be small in signal events. 
$\Delta R_{\gamma\gamma}$  in background process would be quite random because they mainly come from charged particle radiation. 
The Higgs in background process is generally not too boosted, and thus the two photons from the decay of the Higgs tend to go back to back.

\item Invariant mass of the photon pair $m_{\gamma\gamma}$:
\begin{eqnarray}
m_{\gamma\gamma} = \sqrt{( p_{\gamma 1} + p_{\gamma 2} )^2}
\end{eqnarray}
$m_{\gamma\gamma}$ should be around $m_a$ if $\Gamma_a$ is not too large. 
Because the decay constant $f_a$ in Lagrangian~\ref{eq:aLag} is generally larger than TeV~\cite{Cacciapaglia:2019bqz}, so $\Gamma_a < $ 1 GeV can be satisfied in almost all parameter space and we can observe a spiky $m_{\gamma\gamma}$ distribution in signal process.

\item Transverse momentum of visible final state particles also provide useful informations. 
The final state visible particles in signal process come from the decay of heavy $T$ or the recoiling with it, 
while the final state visible particles in background process are from radiation or the decay of $W/h/t$. 
So we can expect that the $p_{\text{T}}$ of the final state visible particles in signal process are larger than the $p_{\text{T}}$ of the final state visible particles in background process. 
Here we consider following variables for discrimination:
\begin{eqnarray}
\nonumber& &\text{$p_{\text{T}}$ of the hardest jet} \ , \ \text{$p_{\text{T}}$ of the second hardest jet} \ , \  \\\nonumber
& &\text{$p_{\text{T}}$ of b-jet} \ , \ \text{$p_{\text{T}}$ of lepton} \ , \ \\\nonumber
& &\text{$p_{\text{T}}$ of the 1st photon} \ , \ \text{$p_{\text{T}}$ of the 2nd photon} 
\end{eqnarray}

\end{itemize}

\begin{figure*}[t!]
\centering
\includegraphics[width=5in]{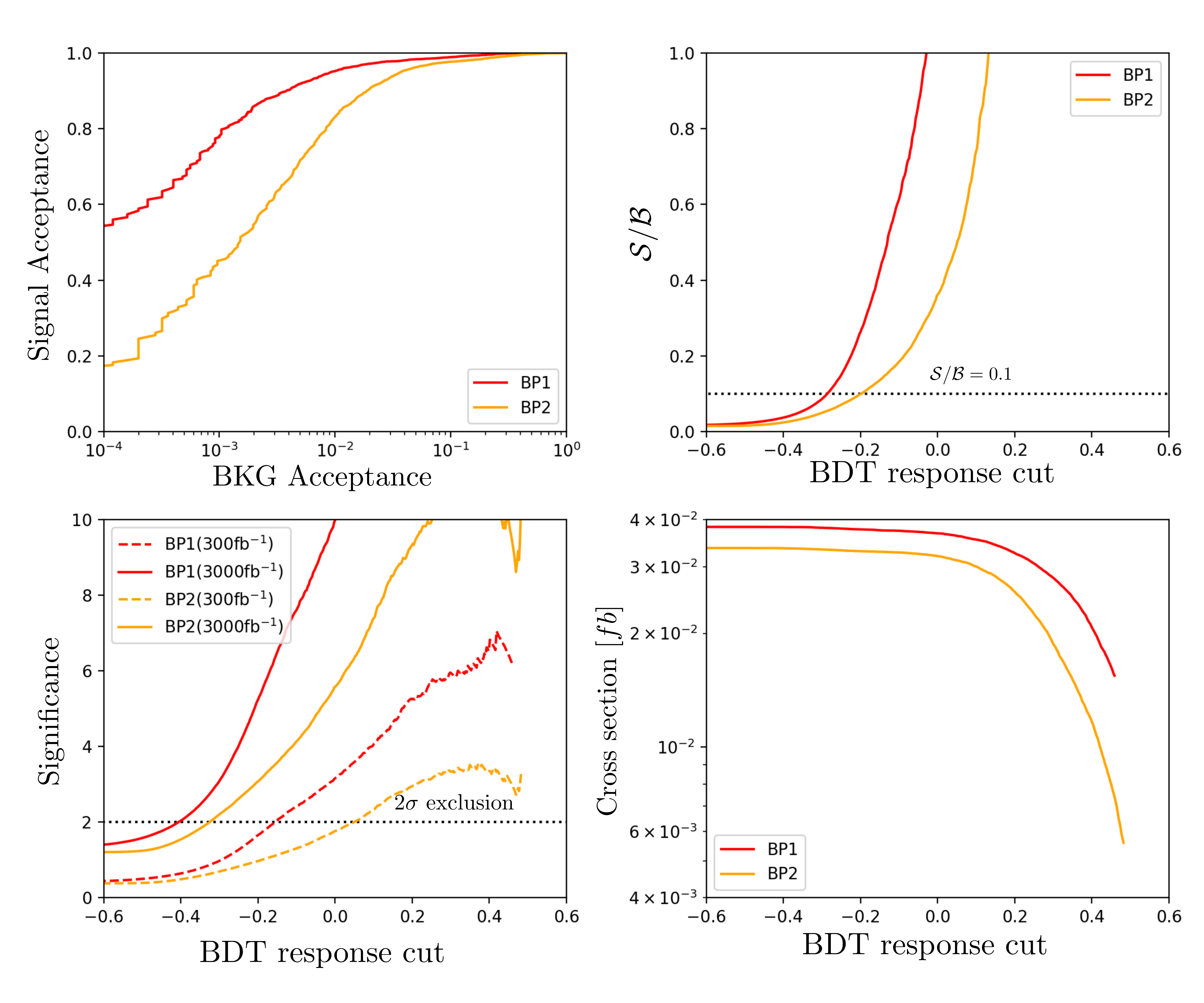}
\caption{({\it up left}): Receiver operating characteristic (ROC) curve for signal and background discrimination. 
({\it up right}): $\mathcal{S}/\mathcal{B}$  as functions of BDT response cut. 
({\it bottom left}): Significance as functions of BDT response cut. Unlike $\mathcal{S}/\mathcal{B}$, Significance also depends on integral luminosity.
({\it bottom right}): Cross section of signal process as functions of BDT response cut.
}
\label{BDT}
\end{figure*}

In Fig.~\ref{variables1} we show the normalized distributions of six variables that are quite different for our BPs and the SM background.
As we expected, for signal process the mass of top partner $T$ and pseudo-scalar $a$ can be reconstructed very well, photon pair are more collinear, and those transverse momentums tend to be harder.
Variables we do not show here also have different distributions for signal and background, but they are subdominant in our signal and background discrimination.

In order to fully utilize all the information to distinguish signal and background, we use Boosted Decision Tree (BDT)~\cite{Roe:2004na}, which is implemented in TMVA-Toolkit~\cite{Hocker:2007ht}, to do a multiple variables analysis.
All the variables after the basic selection are used as input:
\begin{eqnarray}
\left\{
\begin{aligned}
\text{1st jet $\eta$} ,\  \tilde{m}_T ,\ \Delta R_{\gamma\gamma} ,\  m_{\gamma\gamma} ,\ \text{1st jet } p_{\text{T}},\ \\
\text{2nd jet } p_{\text{T}} ,\ \text{1st }\gamma \  p_{\text{T}} ,\ \text{2nd }\gamma \  p_{\text{T}} ,\ l \ p_{\text{T}} , \ \text{b-jet } p_{\text{T}}
\end{aligned}
\right\}.
\end{eqnarray}


In BDT setting, we use 200 decision trees, choose minimum in leaf node as 2.5\%, and set maximum depth as 3. 
Half of the events are chosen as test events, and Kolmogorov-Smirnov test is required to be larger than 0.01 to avoid overtraining. 
BDT maps the multiple variables to a BDT response. 
A signal-like event tends to get a large response, while a background-like event tends to get a small response. 
Thus the cut can be easily performed by requiring the BDT response to be larger than a value\footnote{As present in Fig.~\ref{BDT}, variables' distribution for BP1 and BP2 are different, especially $\Delta R_{\gamma\gamma}$ and $m_{\gamma\gamma}$. So we do not expect that we can use one single BDT to distinguish all model points. Instead, for each $m_T$ and $m_a$, we need to train a corresponding BDT for discrimination. }.
We denote the amount of signal and background events after BDT response cut as $\mathcal{S}$ and $\mathcal{B}$ respectively. 
Statistical significance is evaluated by Poisson formula
\begin{eqnarray}\label{ss}
\text{Significance} =\sqrt{2\left[(\mathcal{S}+\mathcal{B}){\rm ln}(1+\frac{\mathcal{S}}{\mathcal{B}})-\mathcal{S}\right]}.
\end{eqnarray}
If $\text{Significance} > 2$ and the experiment result is consistent with the SM expectation, then this model is excluded at 2$\sigma$ confidence level.
However, sometimes our BDT cut can remove the background very effectively and make the search almost background free ($\mathcal{B} \sim 0$).
In that case we can not use Eq.~\ref{ss} to estimate the statistical significance, but should use $\mathcal{S} = 3$ to determine the 2$\sigma$ exclusion limits\footnote{This is also called ``rule of three" in statistics.}. 
So we require $\mathcal{S}$ to be always greater than 3 to prevent overestimating $\text{Significance}$.  
Considering the systematics at hadron collider, we also require $\mathcal{S}/\mathcal{B}$ to be larger than 0.1.
In Fig.~\ref{BDT}({\it up left}) we present receiver operating characteristic (ROC) curves obtained by BDT response cut. 
It clearly shows that we can lower the background by 3 or 4 orders of magnitude without hurting the signal too much. 
In Fig.~\ref{BDT}({\it up right}), ({\it bottom left}), and ({\it bottom right}) we present $\mathcal{S}/\mathcal{B}$, Significance, and cross section of signal process as functions of BDT response cut. 
It can be seen that we can change the BDT response cut to rapidly increase the $\text{Significance}$ and $\mathcal{S}/\mathcal{B}$. 
And thus $\text{Significance}>2$, $\mathcal{S}/\mathcal{B}>0.1$, and $\mathcal{S} \geqslant 3$ can be satisfied at the same time. 
So both BP1 and BP2 can be excluded by single production process at 14 TeV LHC, no matter the Luminosity is 300 $fb^{-1}$ or 3000 $fb^{-1}$.

\begin{figure*}[ht]
\centering
\includegraphics[width=7in]{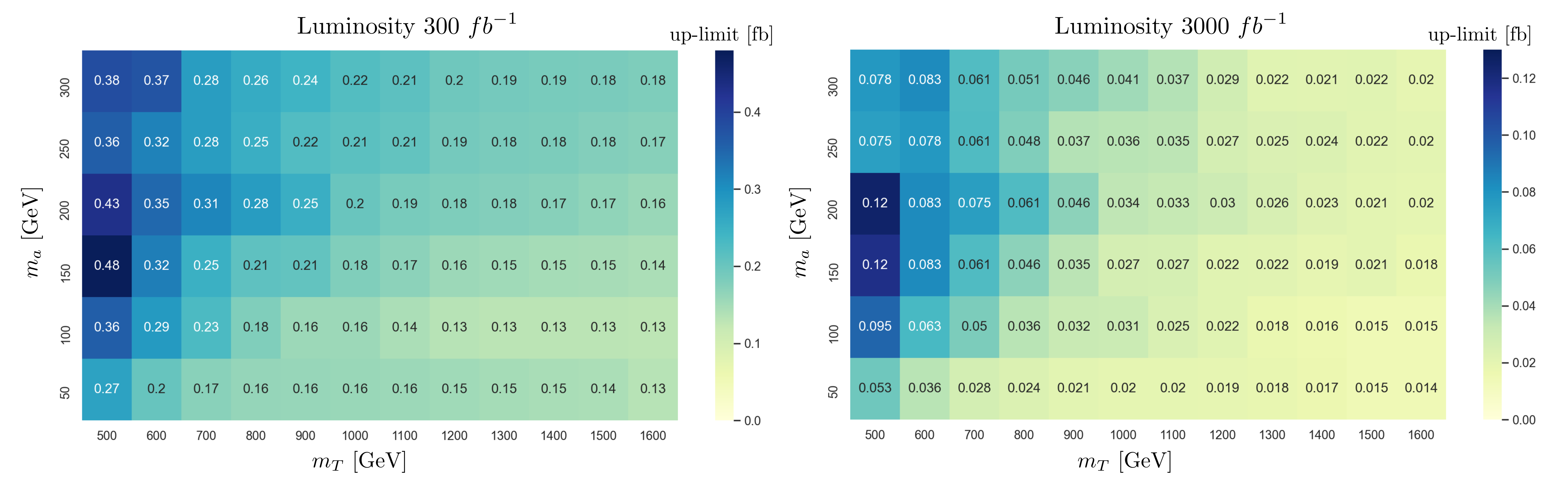}
\caption{
Model independent 2$\sigma$ exclusion limit on  $\sigma(pp\to Tj)\times BR(T\to at)\times BR(a\to\gamma\gamma)$ as a function of $(m_T,\ m_a)$, with integral luminosity 300 fb$^{-1}$ ({\it left}) and 3000 fb$^{-1}$ ({\it right}).
}
\label{plot1}
\end{figure*}

Inversely, if we treat production cross section and BRs as free parameters, then the distributions of those variables can be used to determine the 2$\sigma$ exclusion up limit on
$\sigma(pp\to Tj)\times BR(T\to at)\times BR(a\to\gamma\gamma)$, as a function of $(m_T,\ m_a)$.
For certain $(m_T,\ m_a)$, up limit of $\sigma(pp\to Tj)\times BR(T\to at)\times BR(a\to\gamma\gamma)$ is its minimum value that satisfy 
$\text{Significance}>2$, $\mathcal{S}/\mathcal{B}>0.1$, and $\mathcal{S} \geqslant 3$, under the optimal BDT response cut.  
Repeating this process for each $(m_T,\ m_a)$, we obtain Fig.~\ref{plot1}, the up-limits of $\sigma(pp\to Tj)\times BR(T\to at)\times BR(a\to\gamma\gamma)$ on $m_a$ V.S. $m_T$ plane.
Up-limit with luminosity 3000 fb$^{-1}$ is certainly much smaller than the up-limit with luminosity 300 fb$^{-1}$, 
because Significance is approximately proportional to the square root of integral luminosity.
Points with lower $m_a$ are easier to be excluded. This trend can be understood by the ROC curves in Fig.~\ref{BDT}, the ROC curve for BP1 is on the top of the ROC curve for BP2.
Besides, up limit on $\sigma(pp\to Tj)\times BR(T\to at)\times BR(a\to\gamma\gamma)$ decreases as $m_T$ increases.
It is simply because the decay products of $T$ become more and more energetic as $m_T$ increases, and thus it will be easier to distinguish the signal from the SM background. 

The up-limits we obtained here can be interpreted to a concrete model, by simply calculating $\sigma(pp\to Tj)$ and branching ratios. 
We will show it in Sec.~\ref{sec4}.

\subsection{pair production of top partner}
For pair production process, due to a small $BR(a\to\gamma\gamma)$, we only require one photon pair in the final state. 
The full process is $pp\to T\bar{T}aa\to t\bar{t}\gamma\gamma gg$, followed by $t$ decaying leptonically. 
Like what we have done for single production process, we also use ``basic selection + BDT'' to study pair production process. 
The basic selection used for pair production are:
\begin{itemize}
\item[1.] Exactly two b-jets with $p_\text{T} > 20$ GeV and $|\eta| < 2.5$.
\item[2.] Exactly two leptons (electron or muon) with $p_\text{T} > 20$ GeV and $|\eta| < 2.5$.
\item[3.] Exactly two photons with $p_\text{T} > 20$ GeV and $|\eta| < 2.5$.
\end{itemize}
Here we simply require the final state particles that can be used to suppress the QCD background to exist.

\begin{table*}[t!]
\begin{center}\begin{tabular}{|c|c|c|c|c|c|c|c|c|}
\hline  Basic Selection & \tabincell{c}{BP1 \\ (fb)} & \tabincell{c}{BP2 \\ (fb)} & \tabincell{c}{$t\bar{t}h$\\$(h\to\gamma\gamma)$ \\ (fb)} & \tabincell{c}{ $Whjj$\\$(h\to\gamma\gamma)$ \\ (fb)} & \tabincell{c}{ $t\bar{t}\gamma\gamma$ \\ (fb)}  &\tabincell{c}{ $tj\gamma\gamma$ \\ (fb)} & \tabincell{c}{ $Wjj\gamma\gamma$ \\ (fb)} & \tabincell{c}{ $thj$\\$(h\to\gamma\gamma)$ \\ (fb) }   \\
\hline \tabincell{c}{2 b-jet with \\ $p_\text{T} > 20$ GeV \\ and $|\eta| < 2.5$} &0.656 &0.881 & 0.179 & 0.00327 & 2.39 & 1.38 & 1.46 & 0.00698   \\
\hline \tabincell{c}{2 lepton with \\ $p_\text{T} > 20$ GeV \\ and $|\eta| < 2.5$}&0.090 &0.105 & 0.0133 & 0.00009 & 0.172 & 0.02107 & 0.0132 & 0.00030 \\
\hline  \tabincell{c}{2 $\gamma$ with \\ $p_\text{T} > 20$ GeV \\ and $|\eta| < 2.5$} &0.0295 &0.0328 & 0.00821 & 0.00005 &0.0615 & 0.00596 & 0.00602 & 0.00020   \\
\hline \end{tabular} \caption{Cut-flow table of our basic selection criteria for pair production process.   }
\label{Cutflow2}
\end{center}
\end{table*}

\begin{figure*}[ht]
\centering
\includegraphics[width=6.2in]{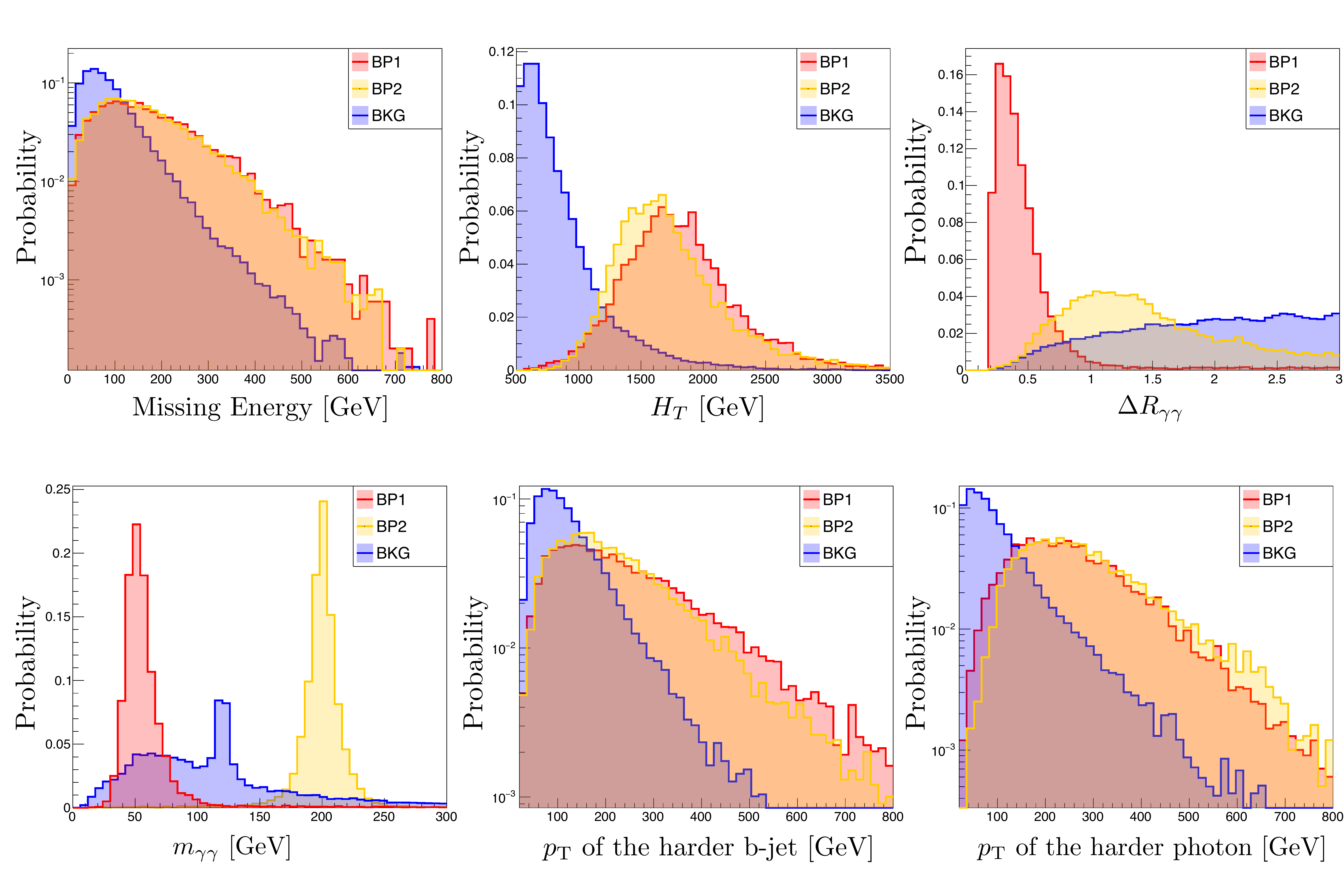}
\caption{Normalized distributions of the variables that we used to distinguish signal from background. Here BKG (background) means the combination of all the six background process. The proportion of each process is proportional to their cross-section after the basic selection.}
\label{variables2}
\end{figure*}

\begin{figure*}[ht]
\centering
\includegraphics[width=5in]{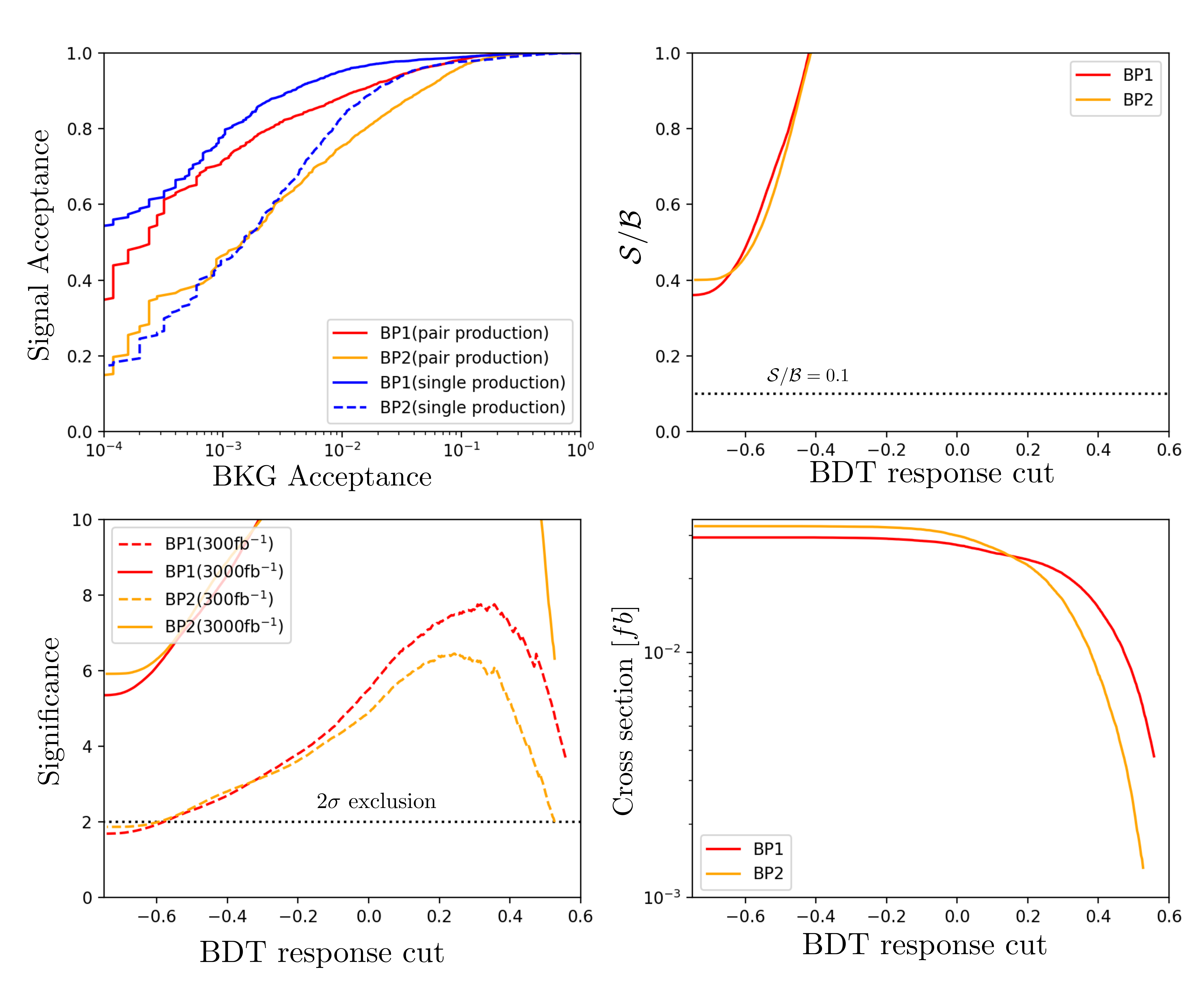}
\caption{({\it up left}): Receiver operating characteristic (ROC) curve for signal and background discrimination. 
({\it up right}): $\mathcal{S}/\mathcal{B}$  as functions of BDT response cut. 
({\it bottom left}): Significance as functions of BDT response cut. 
({\it bottom right}): Cross section of signal process as functions of BDT response cut.
}
\label{BDT2}
\end{figure*}

\begin{figure*}[ht]
\centering
\includegraphics[width=7in]{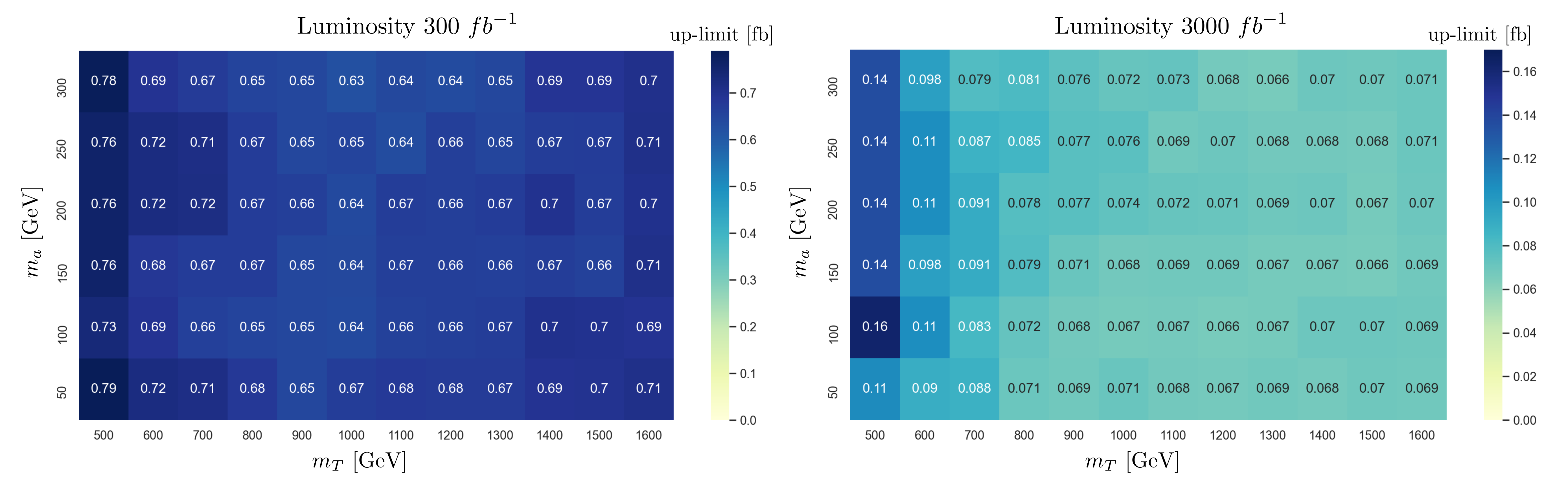}
\caption{
Model independent 2$\sigma$ exclusion limit on  $\sigma(pp\to \bar{T}T)\times \left[BR(T\to at)\right]^2 \times BR(a\to\gamma\gamma)$ as a function of $(m_T,\ m_a)$, with integral luminosity 300 fb$^{-1}$ ({\it left}) and 3000 fb$^{-1}$ ({\it right}).
}
\label{plot2}
\end{figure*}

For illustration, we use the same benchmark points as we used in the last sub-section:
\begin{eqnarray}
\text{BP 1: } m_T &=& 800\text{ GeV} \ , \ m_a = 50\text{ GeV} \\ 
\text{BP 2: } m_T &=& 800\text{ GeV} \ , \ m_a = 200\text{ GeV} 
\end{eqnarray}
And we also fix $BR(T\to ta)$ and $BR(a\to\gamma\gamma)$  to 100\% and 1\% to estimate the search sensitivity of these two BPs. 
The value of $\kappa^T_W$ is not important here, because we assume $BR(T\to bW)$ to be negligible. 
The cut-flow table of our basic selection for these two BPs and the main backgrounds is given in Tab.~\ref{Cutflow2}. 
Because we require two b-jets and two leptons in the final state, so compared with the result present in Tab.~\ref{Cutflow1}, cross section for BKG processes after the  basic selection are much smaller in Tab.~\ref{Cutflow2}.

\begin{figure*}[ht!]
\centering
\includegraphics[width=7in]{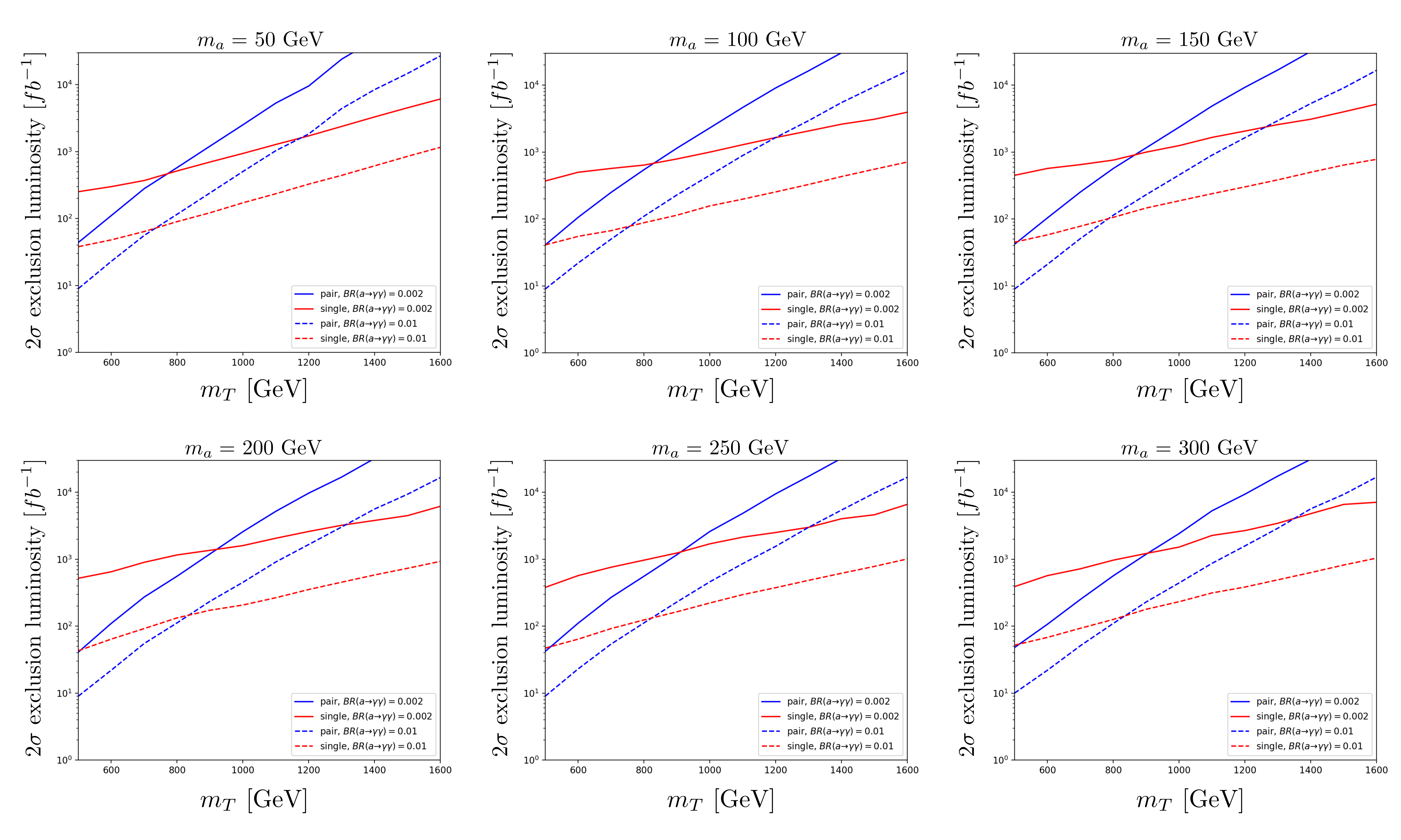}
\caption{
2$\sigma$ exclusion integral luminosity as functions of $m_T$ for single and pair production at 14 TeV LHC. Several values for $m_a$ and $BR(a\to\gamma\gamma)$ are chosen. $BR(T\to ta)$ and $\kappa^T_{W}$ are fixed to 1 and 0.1 respectively.
}
\label{compare1}
\end{figure*}

In our pair process, there are two branches of particles coming from the decay of $T\bar{T}$, so it is difficult for us to distinguish which final state particles originate from the same mother particle. 
And thus it is not easy to reconstruct the mass of top partner.
Instead, we replace reconstructed top partner mass $\tilde{m}_T$ by the scalar $p_{T}$ sum of all final state objects and missing energy, $H_{T}$:
\begin{eqnarray}
H_{T} = \sum_i p_\text{T,i} + {E\!\!\!/}_{\text{T}}
\end{eqnarray}
Here index "$i$" denote two selected b-jets, two selected leptons, two selected photons, and the first and second hardest jet. 
$H_{T}$ roughly reflects the energy scale of hard process, and helps to distinguish signal and background. 
Furthermore, we replace $\eta$ of the hardest jet by missing energy ${E\!\!\!/}_{\text{T}}$, because of the two heavy resonances $T$, generally speaking ${E\!\!\!/}_{\text{T}}$ in signal process is larger than ${E\!\!\!/}_{\text{T}}$ in the SM background process. 

For pair production process, we use following variables as input of BDT to distinguish signal and background:
\begin{eqnarray}
\left\{
\begin{aligned}
{E\!\!\!/}_{\text{T}} ,\  H_{T} ,\ \Delta R_{\gamma\gamma} ,\  m_{\gamma\gamma} ,\ \text{1st b-jet } p_{\text{T}}
,\ \text{2nd b-jet } p_{\text{T}} ,\ \\
\text{1st }\gamma \  p_{\text{T}} ,\ \text{2nd }\gamma \  p_{\text{T}} ,\ \text{1st lepton } p_{\text{T}}
,\ \text{2nd lepton } p_{\text{T}}
\end{aligned}
\right\}. \nonumber
\end{eqnarray}

In Fig.~\ref{variables2} we show the normalized distributions of six variables that are very different for our benchmark signal settings and the SM backgrounds.
As we expected,  the distribution of $H_{T}$ picks around $2m_T$ for signal process. 
The BDT setting we used here is the same as we used in the last sub-section. 
As we did before, in Fig.~\ref{BDT2} we present ROC curves, $\mathcal{S}/\mathcal{B}$, Significance, and cross section of signal process as functions of BDT response cut. 
To compare the search sensitivity of single production and pair production, in Fig.~\ref{BDT2}({\it up left}) we also show the ROC curves obtained from single production channel. 
It can be seen that the area under the ROC curves from single production process is larger than the area under the ROC curves from pair production process. 
While the $\mathcal{S}/\mathcal{B}$ and Significance curves indicate that our BPs can be excluded (or detected) with a much larger Significance through the pair production process\footnote{However, the Significance curve is calculated by Eq.~\ref{ss}, and it does not necessarily mean that the pair production process has a better exclusion ability when $m_T=800$GeV. 
When we decrease $BR(a\to\gamma\gamma)$ to a lower order, we need to reduce $\mathcal{B}$ to close to zero.
In this case, the 2$\sigma$ exclusion limits should be obtained from $\mathcal{S}=3$.}. 
This is simply because the BKG cross section become much smaller, as we have shown in Tab.~\ref{Cutflow2}.  

Like what we did for single production process, we also present the 2$\sigma$ exclusion up limit on $\sigma(pp\to \bar{T}T)\times \left[BR(T\to at)\right]^2 \times BR(a\to\gamma\gamma)$ as a function of $(m_T,\ m_a)$. Fig.~\ref{plot2} is our result. 
The comparison of the exclusion ability of single and pair production can not be directly obtained from Fig.~\ref{plot1} and Fig.~\ref{plot2},
because these two processes have different cross sections. 
In the next sub-section we use more clear and understandable plots to do a comparison.

\subsection{single production V.S. pair production}

Another issue we try to study in this work is the comparison of search sensitivity between single production and the conventional pair production. 
As illustrated in Fig.~\ref{cx}, $\sigma(pp\to Tj)$ will excess $\sigma(pp\to T\bar{T})$ when $m_T$ is large enough. 
But for a realistic analysis, we also need to consider the effect of background. 
It is obvious from Fig.~\ref{plot1} and Fig.~\ref{plot2} that the exclusion up-limits of single production are much smaller than the exclusion up-limits of pair production. But the production crosssections of these two processes are also different. 
In order to clearly compare the search sensitivity, it is better to show the minimal integral luminosity at 14 TeV LHC that is needed for a parameter point to be excluded at 2$\sigma$ level.
For this purpose, we fix $BR(T\to ta)=1$ and $\kappa^T_{W} = 0.1$, then choose several specific values for $m_a$ and $BR(a\to \gamma\gamma)$. 
Then the 2$\sigma$ exclusion integral luminosity can be present as a function of $m_T$. 
Fig.~\ref{compare1} is the result. It clearly shows that pair production is more sensitive when $m_T\lesssim 750$ GeV (850 GeV), and single production is more sensitive when $m_T\gtrsim 750$ GeV (850 GeV), when $m_a$ is 50 GeV (300 GeV).
The dependence on $m_a$ can be understood by the ROC curves in Fig.~\ref{BDT2} and cut-flow tables Tab.~\ref{Cutflow1} and Tab.~\ref{Cutflow2}.

\begin{figure*}[ht!]
\centering
\includegraphics[width=7in]{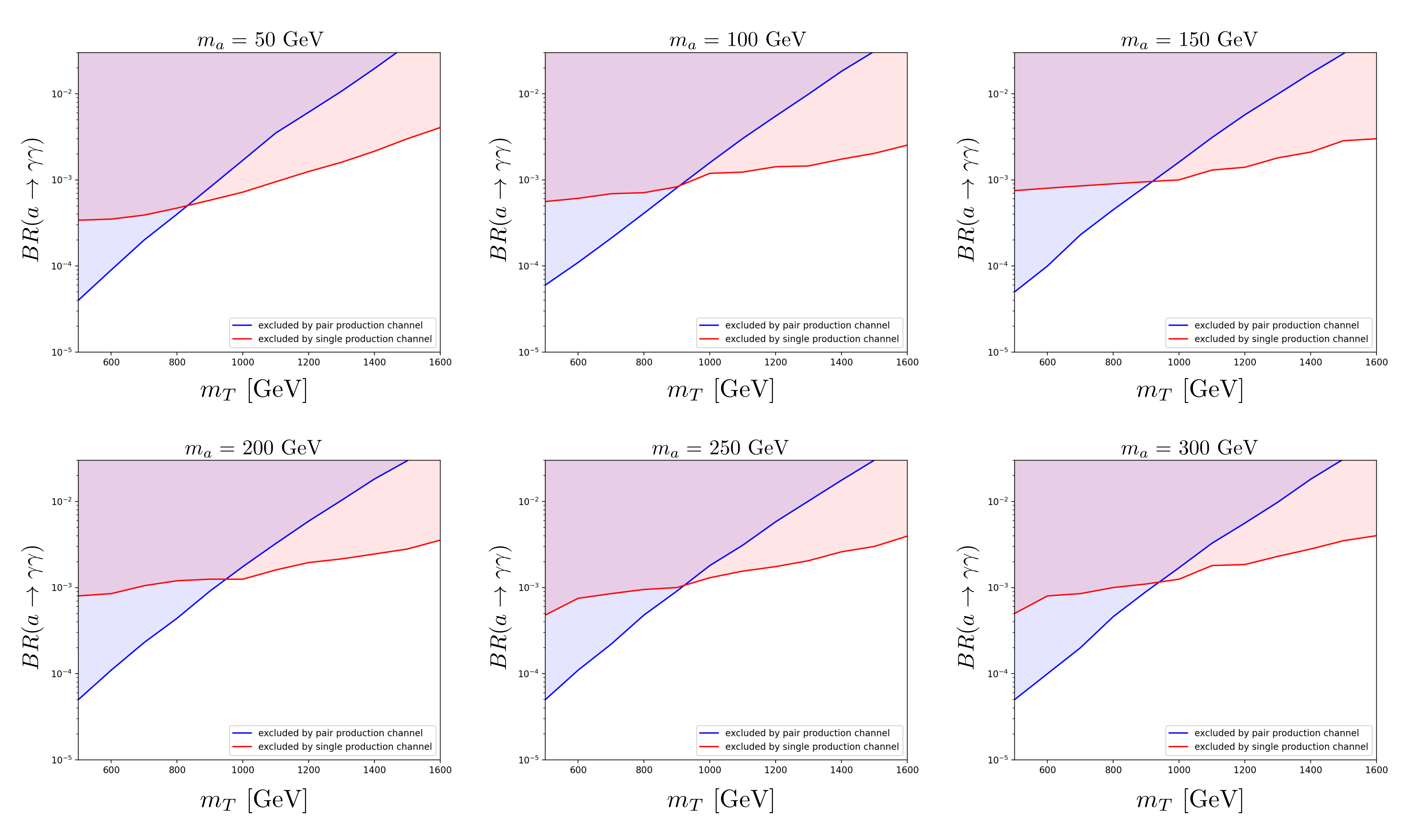}
\caption{
2$\sigma$ exclusion limit on $m_T$ V.S. $BR(a\to\gamma\gamma)$ plane for single and pair production.
Here we consider 14 TeV LHC with integral luminosity 3000 $fb^{-1}$. 
$BR(T\to ta)$ and $\kappa^T_{W}$ are fixed to 1 and 0.1 respectively.
}
\label{compare2}
\end{figure*}

A more realistic comparison can be performed by fixing the 14 TeV LHC integral luminosity to 3000 $fb^{-1}$, and treat $BR(a\to\gamma\gamma)$ as a free parameter. 
Then pair production channel and single production channel can exclude two different regions on $m_T$ V.S. $BR(a\to\gamma\gamma)$ plane at 2$\sigma$ level.
In Fig.~\ref{compare2} we present the excluded region, and it clearly shows that the single production channel excludes more parameter space when $m_T\gtrsim 800$ GeV (900 GeV) and $m_a = 50$ GeV (300 GeV).

\section{Interpretation of concrete model}
\label{sec4}
\begin{figure}[ht!]
\centering
\includegraphics[width=3.3in]{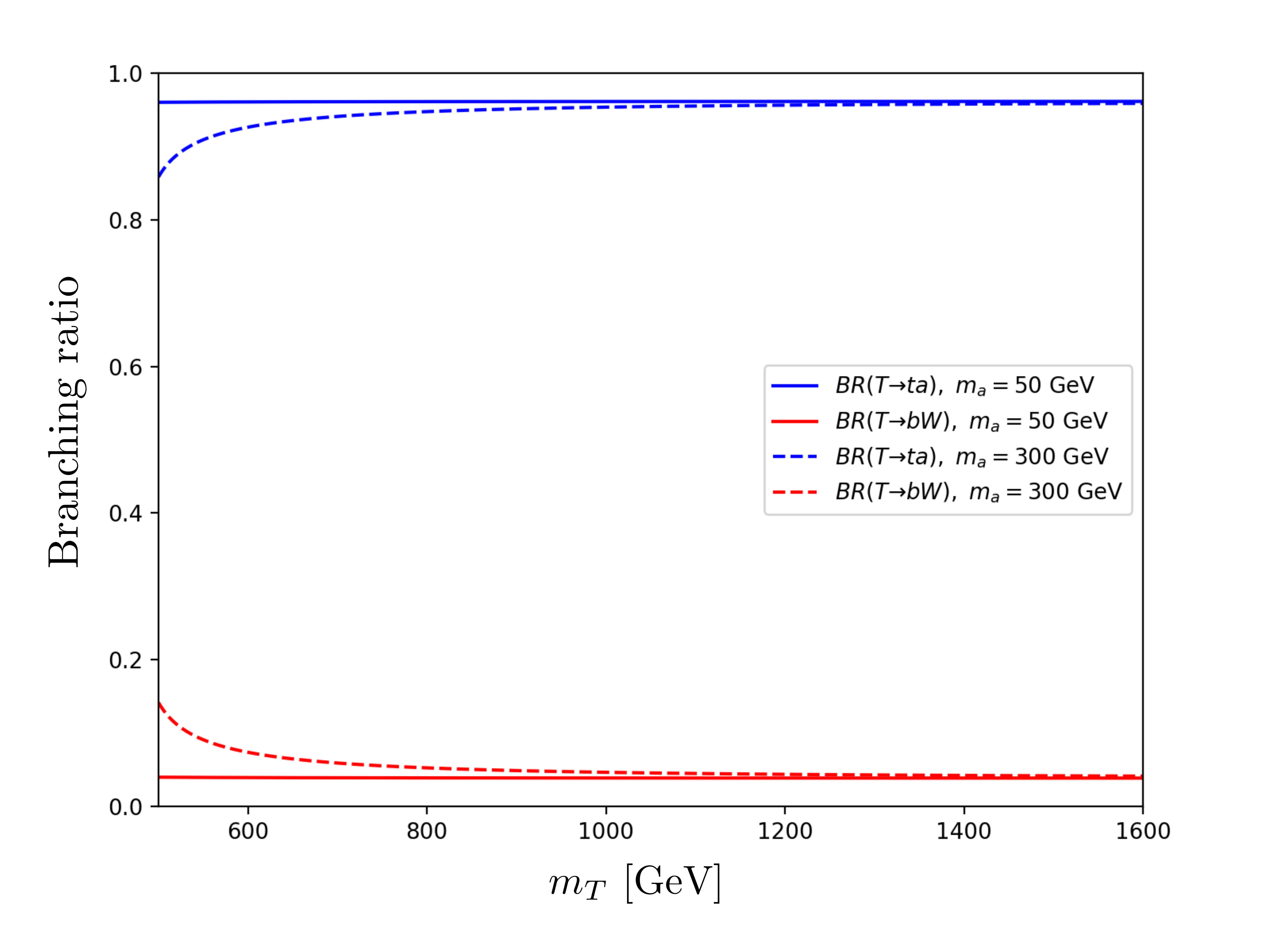}
\caption{
$BR(T\to ta)$ and $BR(T\to bW)$ in Model 1(2) as functions of $m_T$, under different $m_a$ value. 
}
\label{BR}
\end{figure}

In this section we interpret the model independent exclusion limits we obtained in the last section to concrete models. 
As explained in Sec. II, in this work we treat $T\to Zt$, $T\to ht$, and $a\to \bar{f}f$ as negligible.
And for simplicity we further close all the right-hand couplings.
Thus our Lagrangian can be simplified as:
\begin{eqnarray}
\mathcal{L}_{T} &=& \phantom{+}  \overline{T}\left(i\slashed{D}-m_T\right) T    \\\nonumber
 & & + \left(\kappa^T_{W} \frac{g}{\sqrt{2}}  \,\overline{T}\slashed{W}^+P_Lb + i \kappa^T_{a} \frac{m_T}{v} \, \overline{T} a P_L t +  \mbox{ h.c. }\right) \\ 
{\mathcal{L}}_a &=& \frac{1}{2}(\partial_\mu a)(\partial^\mu a) 	-\frac{1}{2} m_a^2 a^2 + \frac{g_s^2 K_g^a }{16\pi^2 f_a}a {G}^a_{\mu\nu}\tilde{{G}}^{a\mu\nu}  \\ \nonumber
 & &+ \frac{e^2 K^a_\gamma}{16\pi^2 f_a} a A_{\mu\nu} \tilde{A}^{\mu\nu} + \frac{g^2 c_W^2 K^a_Z}{16 \pi^2 f_a} a Z_{\mu\nu} \tilde{Z}^{\mu\nu}    \\\nonumber
	& &  + \frac{egc_W K^a_{Z\gamma}}{8\pi^2 f_a} a A_{\mu\nu} \tilde{Z}^{\mu\nu} 
\end{eqnarray}
$\sigma(pp\to Tj)$ is proportional to $(\kappa^T_{W})^2$, and the decay width of $T$ can be expressed as:
\begin{eqnarray}
\Gamma (T \to W b) &=& (\kappa^T_{W})^2 \frac{m_T^3 g^2}{64 \pi m_W^2} \Gamma_{W} (m_T, m_W, m_b) \label{eq:GammaW} \\
\Gamma (T \to t a) &=& (\kappa^T_{a})^2  \frac{m_T^3 g^2}{64 \pi m_W^2} \Gamma_{a} (m_T, m_a, m_t)\label{eq:Gammaa}
\end{eqnarray}
Those kinematic functions are:
\begin{eqnarray}
& &\Gamma_{W}(m_T, m_W, m_b) = \lambda^{\frac{1}{2}} (1, \frac{m_b^2}{m_T^2}, \frac{m_W^2}{m_T^2}) \times \\ \nonumber
& & \ \ \ \ \ \ \ \ \ \ \ \ \ \  \left[ \left( 1-\frac{m_b^2}{m_T^2} \right)^2 + \frac{m_W^2}{m_T^2} - 2 \frac{m_W^4}{m_T^4} + \frac{m_W^2 m_b^2}{m_T^4} \right]  \\\nonumber
& &\Gamma_{a}(m_T, m_a, m_t) = \frac{1}{2} \lambda^{\frac{1}{2}} (1, \frac{m_t^2}{m_T^2}, \frac{m_a^2}{m_T^2})  \left[ 1 + \frac{m_t^2}{m_T^2} -  \frac{m_a^2}{m_T^2} \right]
\end{eqnarray}
with phase space function $\lambda (a, b, c)$  defined as:
\bea
\lambda (a, b, c) = a^2 + b^2 + c^2 - 2 ab - 2 ac - 2 bc
\eea

\begin{figure*}[ht!]
\centering
\includegraphics[width=7in]{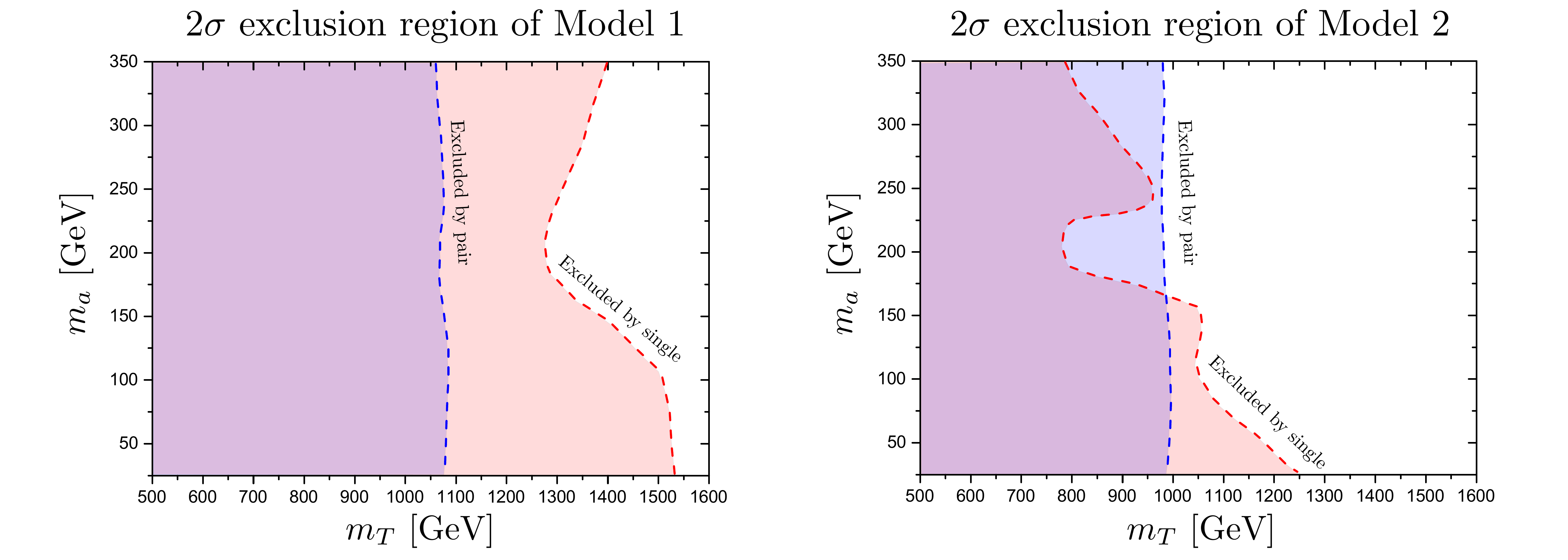}
\caption{
Left: 2$\sigma$ exclusion region on $m_T - m_a$ plane for Model 1. 
Left: 2$\sigma$ exclusion region on $m_T - m_a$ plane for Model 2. 
Here we only show the exclusion limits at $3000$ fb$^{-1}$ 14 TeV LHC.
}
\label{exclu2}
\end{figure*}

For pseudo-scalar $a$, actually we only need to know the ratio between $\Gamma(a\to \gamma\gamma)$ and $\Gamma(a\to gg)$.
It is because strong coupling $g_s$ is much larger than electro-weak coupling $g$, thus the total decay width of $a$ is approximately determined by $\Gamma(a\to gg)$. 
Then $BR(a\to \gamma\gamma)$ can be estimated by:
\begin{eqnarray}
BR(a\to \gamma\gamma) \approx  \frac{\Gamma(a\to \gamma\gamma)}{\Gamma(a\to gg)} = \frac{\alpha^2_{EM}(K^a_\gamma)^2  }{8\alpha^2_S (K^a_g)^2 } 
\end{eqnarray}
Here we treat $(m_T,\ m_a)$ as undetermined parameters. 
$K^a_Z$, $K^a_{Z\gamma}$, $K^a_W$, and $f_a$ are not relevant to our collider analysis.
Limits from di-boson resonance search can be easily avoided by assuming $f_a$ a large value. 
So except $m_T$ and $m_a$, we only have four input parameters for a simplified model:
\begin{eqnarray}
\kappa^T_{W}  \ ,  \    \kappa^T_{a} \ ,  \     K^a_\gamma  \ ,  \   K^a_g
\end{eqnarray}

For illustration, we consider two models for exclusion limits interpreting:
\begin{eqnarray}
& &\text{Model 1: } \nonumber   \\ & &  \kappa^T_{W} = 0.1  \ ,  \    \kappa^T_{a} = 0.5  \ ,  \     K^a_\gamma = 1.0 \ ,  \   K^a_g = 0.5         \\
& &\text{Model 2: } \nonumber   \\  & &   \kappa^T_{W} = 0.1 \ ,  \    \kappa^T_{a} = 0.5  \ ,  \     K^a_\gamma = 1.2 \ ,  \   K^a_g = 0.8
\end{eqnarray}
The difference of these two models is the value of $BR(a\to\gamma\gamma)$, which are 0.22\% and 0.12\% for Model 1 and Model 2.
$BR(T\to ta)$ and $BR(T\to bW)$ are the same for two models, but they depend on $m_a$ and $m_T$. 
In Fig.~\ref{BR} we present $BR(T\to ta)$ and $BR(T\to bW)$ as functions of $m_T$, with $m_a$ fixed to 50 GeV and 300 GeV.
It can be seen that in most of the parameter space, $BR(T\to bW)$ is smaller than 5\%, and thus the current direct search bound can be escaped~\cite{Aaboud:2017zfn}.

After calculating $BR(T\to at)$ and $BR(a\to\gamma\gamma)$ from these couplings and spectrum,
 we can compare $\sigma(pp\to Tj)\times BR(T\to at)\times BR(a\to\gamma\gamma)$ and $\sigma(pp\to T\bar{T})\times \left[BR(T\to at)\right]^2\times BR(a\to\gamma\gamma)$ of each $(m_T,\ m_a)$ with the up limit presented in Fig.~\ref{plot1} and Fig.~\ref{plot2}.
Then the 2$\sigma$ exclusion limit on $m_T - m_a$ plane can be obtained. 
Fig.~\ref{exclu2} is the exclusion limits on $m_T - m_a$ plane for both models. 
It shows that if $BR(T\to ta)\approx1$ and $BR(a\to \gamma\gamma)$ is around $\mathcal{O}(0.1\%)$, we can exclude $m_T$ up to TeV scale at future high luminosity LHC. 
The detecting ability of single production is more sensitive on the value of $BR(a\to \gamma\gamma)$. 
Sensitivity of single production channel is greatly enhanced when $BR(a\to \gamma\gamma)$ increases from 0.12\% to 0.22\%. 
Our results also shows that the single production channel becomes more sensitive when pNGB $a$ is getting lighter. 
On the contrary, pair production search channel is more robust against the value of $BR(a\to \gamma\gamma)$ and $m_a$.


\section{Conclusions}
The vector-like top partner and the pseudo-Nambu Goldstone Boson are two key features of composite Higgs models.  
In this paper, we studied the observability of a new signature of top partner decaying to pNGB, $T\to t a $, through the production processes $pp \to Tj$ and $pp \to T \bar{T}$ at the LHC. 
We found the the clean decay channel of pNGB, $a\to\gamma\gamma$, helps to suppress the huge QCD background, even the branching ratio of $a\to\gamma\gamma$ is as small as $\mathcal{O}(0.1\%)$.
Model independent exclusion limits for single and pair production are present, and can be easily interpreted to concrete model. 
We also compare the direct search sensitivity of single and pair production channel. 
We found that the single production process $pp\to Tj$ can be more sensitive than the conventional pair production process $pp\to \bar{T}T$ when $m_T$ is larger than $800\sim 900$ GeV.

\section*{Acknowledgements}
Part of this work was done while authors were visiting Jinan University. This work was supported by the National Natural Science Foundation of China (NNSFC) under grant Nos. 11705093 and 11947118.


\begin{thebibliography}{99}
\bibitem{Aad:2012tfa} 
  G.~Aad {\it et al.} [ATLAS Collaboration],
  ``Observation of a new particle in the search for the Standard Model Higgs boson with the ATLAS detector at the LHC,''
  Phys.\ Lett.\ B {\bf 716}, 1 (2012)
  doi:10.1016/j.physletb.2012.08.020
  [arXiv:1207.7214 [hep-ex]].

\bibitem{Chatrchyan:2012xdj} 
  S.~Chatrchyan {\it et al.} [CMS Collaboration],
  ``Observation of a New Boson at a Mass of 125 GeV with the CMS Experiment at the LHC,''
  Phys.\ Lett.\ B {\bf 716}, 30 (2012)
  doi:10.1016/j.physletb.2012.08.021
  [arXiv:1207.7235 [hep-ex]].

\bibitem{Kaplan:1983fs} 
  D.~B.~Kaplan and H.~Georgi,
  ``SU(2) x U(1) Breaking by Vacuum Misalignment,''
  Phys.\ Lett.\  {\bf 136B}, 183 (1984).
  doi:10.1016/0370-2693(84)91177-8

\bibitem{Kaplan:1991dc} 
  D.~B.~Kaplan,
  ``Flavor at SSC energies: A New mechanism for dynamically generated fermion masses,''
  Nucl.\ Phys.\ B {\bf 365}, 259 (1991).
  doi:10.1016/S0550-3213(05)80021-5
  
  



\bibitem{Redi:2012ha} 
  M.~Redi and A.~Tesi,
  ``Implications of a Light Higgs in Composite Models,''
  JHEP {\bf 1210}, 166 (2012)
  doi:10.1007/JHEP10(2012)166
  [arXiv:1205.0232 [hep-ph]].
  
\bibitem{Contino:2006qr} 
  R.~Contino, L.~Da Rold and A.~Pomarol,
  ``Light custodians in natural composite Higgs models,''
  Phys.\ Rev.\ D {\bf 75}, 055014 (2007)
  doi:10.1103/PhysRevD.75.055014
  [hep-ph/0612048].
  
\bibitem{Matsedonskyi:2012ym} 
  O.~Matsedonskyi, G.~Panico and A.~Wulzer,
  ``Light Top Partners for a Light Composite Higgs,''
  JHEP {\bf 1301}, 164 (2013)
  doi:10.1007/JHEP01(2013)164
  [arXiv:1204.6333 [hep-ph]].

\bibitem{Marzocca:2012zn} 
  D.~Marzocca, M.~Serone and J.~Shu,
  ``General Composite Higgs Models,''
  JHEP {\bf 1208}, 013 (2012)
  doi:10.1007/JHEP08(2012)013
  [arXiv:1205.0770 [hep-ph]].

\bibitem{Blasi:2019jqc}
S.~Blasi and F.~Goertz,
``Softened Symmetry Breaking in Composite Higgs Models,''
Phys. Rev. Lett. \textbf{123}, no.22, 221801 (2019)
doi:10.1103/PhysRevLett.123.221801
[arXiv:1903.06146 [hep-ph]].

\bibitem{Blasi:2020ktl}
S.~Blasi, C.~Csaki and F.~Goertz,
``A Natural Composite Higgs via Universal Boundary Conditions,''
[arXiv:2004.06120 [hep-ph]].


\bibitem{Agashe:2004rs} 
  K.~Agashe, R.~Contino and A.~Pomarol,
  ``The Minimal composite Higgs model,''
  Nucl.\ Phys.\ B {\bf 719}, 165 (2005)
  doi:10.1016/j.nuclphysb.2005.04.035
  [hep-ph/0412089].

\bibitem{simone2012partner} 
 A.~D.~Simone, O.~Matsedonskyi, R.~Rattazzi and A.~Wulzer,
 ``A First Top Partner Hunter's Guide,''
 JHEP {\bf 1304}, 004 (2013)
  doi:10.1007/jhep04(2013)004
  [hep-ph/1211.5663].



\bibitem{Aaboud:2018xuw} 
  M.~Aaboud {\it et al.} [ATLAS Collaboration],
  ``Search for pair production of up-type vector-like quarks and for four-top-quark events in final states with multiple $b$-jets with the ATLAS detector,''
  JHEP {\bf 1807}, 089 (2018)
  doi:10.1007/JHEP07(2018)089
  [arXiv:1803.09678 [hep-ex]].

\bibitem{Aaboud:2018saj} 
  M.~Aaboud {\it et al.} [ATLAS Collaboration],
  ``Search for pair- and single-production of vector-like quarks in final states with at least one $Z$ boson decaying into a pair of electrons or muons in $pp$ collision data collected with the ATLAS detector at $\sqrt{s} = 13$ TeV,''
  Phys.\ Rev.\ D {\bf 98}, no. 11, 112010 (2018)
  doi:10.1103/PhysRevD.98.112010
  [arXiv:1806.10555 [hep-ex]].

\bibitem{Aaboud:2017qpr} 
  M.~Aaboud {\it et al.} [ATLAS Collaboration],
  ``Search for pair production of vector-like top quarks in events with one lepton, jets, and missing transverse momentum in $ \sqrt{s}=13 $ TeV $pp$ collisions with the ATLAS detector,''
  JHEP {\bf 1708}, 052 (2017)
  doi:10.1007/JHEP08(2017)052
  [arXiv:1705.10751 [hep-ex]].

\bibitem{Aaboud:2017zfn} 
  M.~Aaboud {\it et al.} [ATLAS Collaboration],
  ``Search for pair production of heavy vector-like quarks decaying to high-p$_{T}$ W bosons and b quarks in the lepton-plus-jets final state in pp collisions at $ \sqrt{s}=13 $ TeV with the ATLAS detector,''
  JHEP {\bf 1710}, 141 (2017)
  doi:10.1007/JHEP10(2017)141
  [arXiv:1707.03347 [hep-ex]].

\bibitem{Aaboud:2018wxv} 
  M.~Aaboud {\it et al.} [ATLAS Collaboration],
  ``Search for pair production of heavy vector-like quarks decaying into hadronic final states in $pp$ collisions at $\sqrt{s} = 13$ TeV with the ATLAS detector,''
  Phys.\ Rev.\ D {\bf 98}, no. 9, 092005 (2018)
  doi:10.1103/PhysRevD.98.092005
  [arXiv:1808.01771 [hep-ex]].

\bibitem{Aaboud:2018uek} 
  M.~Aaboud {\it et al.} [ATLAS Collaboration],
  ``Search for pair production of heavy vector-like quarks decaying into high-$p_T$ $W$ bosons and top quarks in the lepton-plus-jets final state in $pp$ collisions at $\sqrt{s}=13$ TeV with the ATLAS detector,''
  JHEP {\bf 1808}, 048 (2018)
  doi:10.1007/JHEP08(2018)048
  [arXiv:1806.01762 [hep-ex]].

\bibitem{Aaboud:2018pii} 
  M.~Aaboud {\it et al.} [ATLAS Collaboration],
  ``Combination of the searches for pair-produced vector-like partners of the third-generation quarks at $\sqrt{s} =$ 13 TeV with the ATLAS detector,''
  Phys.\ Rev.\ Lett.\  {\bf 121}, no. 21, 211801 (2018)
  doi:10.1103/PhysRevLett.121.211801
  [arXiv:1808.02343 [hep-ex]].

\bibitem{Sirunyan:2018omb} 
  A.~M.~Sirunyan {\it et al.} [CMS Collaboration],
  ``Search for vector-like T and B quark pairs in final states with leptons at $\sqrt{s} =$ 13 TeV,''
  JHEP {\bf 1808}, 177 (2018)
  doi:10.1007/JHEP08(2018)177
  [arXiv:1805.04758 [hep-ex]].

\bibitem{Sirunyan:2017pks} 
  A.~M.~Sirunyan {\it et al.} [CMS Collaboration],
  ``Search for pair production of vector-like quarks in the bW$\overline{\mathrm{b}}$W channel from proton-proton collisions at $\sqrt{s} =$ 13 TeV,''
  Phys.\ Lett.\ B {\bf 779}, 82 (2018)
  doi:10.1016/j.physletb.2018.01.077
  [arXiv:1710.01539 [hep-ex]].

\bibitem{Sirunyan:2019sza} 
  A.~M.~Sirunyan {\it et al.} [CMS Collaboration],
  ``Search for pair production of vectorlike quarks in the fully hadronic final state,''
  Phys.\ Rev.\ D {\bf 100}, no. 7, 072001 (2019)
  doi:10.1103/PhysRevD.100.072001
  [arXiv:1906.11903 [hep-ex]].

\bibitem{Han:2014qia}
C.~Han, A.~Kobakhidze, N.~Liu, L.~Wu and B.~Yang,
``Constraining Top partner and Naturalness at the LHC and TLEP,''
Nucl. Phys. B \textbf{890}, 388-399 (2014)
doi:10.1016/j.nuclphysb.2014.11.021
[arXiv:1405.1498 [hep-ph]].

\bibitem{Liu:2015kmo}
N.~Liu, L.~Wu, B.~Yang and M.~Zhang,
``Single top partner production in the Higgs to diphoton channel in the Littlest Higgs Model with $T$-parity,''
Phys. Lett. B \textbf{753}, 664-669 (2016)
doi:10.1016/j.physletb.2015.12.066
[arXiv:1508.07116 [hep-ph]].


\bibitem{Anandakrishnan:2015yfa} 
  A.~Anandakrishnan, J.~H.~Collins, M.~Farina, E.~Kuflik and M.~Perelstein,
  ``Odd Top Partners at the LHC,''
  Phys.\ Rev.\ D {\bf 93}, no. 7, 075009 (2016)
  doi:10.1103/PhysRevD.93.075009
  [arXiv:1506.05130 [hep-ph]].

\bibitem{Kraml:2016eti} 
  S.~Kraml, U.~Laa, L.~Panizzi and H.~Prager,
  ``Scalar versus fermionic top partner interpretations of $t\bar t + E_T^{\rm miss}$ searches at the LHC,''
  JHEP {\bf 1611}, 107 (2016)
  doi:10.1007/JHEP11(2016)107
  [arXiv:1607.02050 [hep-ph]].

\bibitem{Bizot:2018tds} 
  N.~Bizot, G.~Cacciapaglia and T.~Flacke,
  ``Common exotic decays of top partners,''
  JHEP {\bf 1806}, 065 (2018)
  doi:10.1007/JHEP06(2018)065
  [arXiv:1803.00021 [hep-ph]].

\bibitem{Serra:2015xfa} 
  J.~Serra,
  ``Beyond the Minimal Top Partner Decay,''
  JHEP {\bf 1509}, 176 (2015)
  doi:10.1007/JHEP09(2015)176
  [arXiv:1506.05110 [hep-ph]].

\bibitem{Han:2018hcu} 
  H.~Han, L.~Huang, T.~Ma, J.~Shu, T.~M.~P.~Tait and Y.~Wu,
  ``Six Top Messages of New Physics at the LHC,''
  JHEP {\bf 1910}, 008 (2019)
  doi:10.1007/JHEP10(2019)008
  [arXiv:1812.11286 [hep-ph]].

\bibitem{Alhazmi:2018whk} 
  H.~Alhazmi, J.~H.~Kim, K.~Kong and I.~M.~Lewis,
  ``Shedding Light on Top Partner at the LHC,''
  JHEP {\bf 1901}, 139 (2019)
  doi:10.1007/JHEP01(2019)139
  [arXiv:1808.03649 [hep-ph]].

\bibitem{Benbrik:2019zdp}
R.~Benbrik, E.~B.~Kuutmann, D.~Buarque Franzosi, V.~Ellajosyula, R.~Enberg, G.~Ferretti, M.~Isacson, Y.~B.~Liu, T.~Mandal and T.~Mathisen, \textit{et al.}
``Signatures of vector-like top partners decaying into new neutral scalar or pseudoscalar bosons,''
JHEP \textbf{05}, 028 (2020)
doi:10.1007/JHEP05(2020)028
[arXiv:1907.05929 [hep-ph]].

\bibitem{Aguilar-Saavedra:2017giu} 
  J.~A.~Aguilar-Saavedra, D.~E.~López-Fogliani and C.~Muñoz,
  ``Novel signatures for vector-like quarks,''
  JHEP {\bf 1706}, 095 (2017)
  doi:10.1007/JHEP06(2017)095
  [arXiv:1705.02526 [hep-ph]].

\bibitem{Chala:2017xgc} 
  M.~Chala,
  ``Direct bounds on heavy toplike quarks with standard and exotic decays,''
  Phys.\ Rev.\ D {\bf 96}, no. 1, 015028 (2017)
  doi:10.1103/PhysRevD.96.015028
  [arXiv:1705.03013 [hep-ph]].

\bibitem{Cacciapaglia:2019zmj} 
  G.~Cacciapaglia, T.~Flacke, M.~Park and M.~Zhang,
  ``Exotic decays of top partners: mind the search gap,''
  Phys.\ Lett.\ B {\bf 798}, 135015 (2019)
  doi:10.1016/j.physletb.2019.135015
  [arXiv:1908.07524 [hep-ph]].

\bibitem{Kim:2018mks} 
  J.~H.~Kim and I.~M.~Lewis,
  ``Loop Induced Single Top Partner Production and Decay at the LHC,''
  JHEP {\bf 1805}, 095 (2018)
  doi:10.1007/JHEP05(2018)095
  [arXiv:1803.06351 [hep-ph]].

\bibitem{Xie:2019gya} 
  K.~P.~Xie, G.~Cacciapaglia and T.~Flacke,
  ``Exotic decays of top partners with charge 5/3: bounds and opportunities,''
  JHEP {\bf 1910}, 134 (2019)
  doi:10.1007/JHEP10(2019)134
  [arXiv:1907.05894 [hep-ph]].

\bibitem{Criado:2019mvu} 
  J.~C.~Criado and M.~Perez-Victoria,
  ``Vector-like quarks with non-renormalizable interactions,''
  JHEP {\bf 2001}, 057 (2020)
  doi:10.1007/JHEP01(2020)057
  [arXiv:1908.08964 [hep-ph]].

\bibitem{Aguilar-Saavedra:2019ghg} 
  J.~A.~Aguilar-Saavedra, J.~Alonso-González, L.~Merlo and J.~M.~No,
  ``Exotic vectorlike quark phenomenology in the minimal linear $\sigma$ model,''
  Phys.\ Rev.\ D {\bf 101}, no. 3, 035015 (2020)
  doi:10.1103/PhysRevD.101.035015
  [arXiv:1911.10202 [hep-ph]].

\bibitem{Ramos:2019qqa} 
  M.~Ramos,
  ``Composite dark matter phenomenology in the presence of lighter degrees of freedom,''
  arXiv:1912.11061 [hep-ph].




\bibitem{Barnard:2013zea} 
  J.~Barnard, T.~Gherghetta and T.~S.~Ray,
  JHEP {\bf 1402}, 002 (2014)
  doi:10.1007/JHEP02(2014)002
  [arXiv:1311.6562 [hep-ph]].

\bibitem{Ferretti:2013kya} 
  G.~Ferretti and D.~Karateev,
  JHEP {\bf 1403}, 077 (2014)
  doi:10.1007/JHEP03(2014)077
  [arXiv:1312.5330 [hep-ph]].

\bibitem{Ferretti:2016upr} 
  G.~Ferretti,
  ``Gauge theories of Partial Compositeness: Scenarios for Run-II of the LHC,''
  JHEP {\bf 1606}, 107 (2016)
  doi:10.1007/JHEP06(2016)107
  [arXiv:1604.06467 [hep-ph]].

\bibitem{DeGrand:2016pgq} 
  T.~DeGrand, M.~Golterman, E.~T.~Neil and Y.~Shamir,
  Phys.\ Rev.\ D {\bf 94}, no. 2, 025020 (2016)
  doi:10.1103/PhysRevD.94.025020
  [arXiv:1605.07738 [hep-ph]].



\bibitem{Belyaev:2016ftv} 
  A.~Belyaev, G.~Cacciapaglia, H.~Cai, G.~Ferretti, T.~Flacke, A.~Parolini and H.~Serodio,
  ``Di-boson signatures as Standard Candles for Partial Compositeness,''
  JHEP {\bf 1701}, 094 (2017)
  Erratum: [JHEP {\bf 1712}, 088 (2017)]
  doi:10.1007/JHEP01(2017)094, 10.1007/JHEP12(2017)088
  [arXiv:1610.06591 [hep-ph]].
  
\bibitem{Cacciapaglia:2019bqz} 
  G.~Cacciapaglia, G.~Ferretti, T.~Flacke and H.~Serôdio,
  ``Light scalars in composite Higgs models,''
  Front.\ in Phys.\  {\bf 7}, 22 (2019)
  doi:10.3389/fphy.2019.00022
  [arXiv:1902.06890 [hep-ph]].






\bibitem{Contino:2004vy}
R.~Contino and A.~Pomarol,
``Holography for fermions,''
JHEP \textbf{11}, 058 (2004)
doi:10.1088/1126-6708/2004/11/058
[arXiv:hep-th/0406257 [hep-th]].

\bibitem{Cacciapaglia:2008bi}
G.~Cacciapaglia, G.~Marandella and J.~Terning,
``Dimensions of Supersymmetric Operators from AdS/CFT,''
JHEP \textbf{06}, 027 (2009)
doi:10.1088/1126-6708/2009/06/027
[arXiv:0802.2946 [hep-th]].

\bibitem{Schmaltz:2002wx}
M.~Schmaltz,
``Physics beyond the standard model (theory): Introducing the little Higgs,''
Nucl. Phys. B Proc. Suppl. \textbf{117}, 40-49 (2003)
doi:10.1016/S0920-5632(03)01409-9
[arXiv:hep-ph/0210415 [hep-ph]].



\bibitem{Dermisek:2019vkc}
R.~Dermíšek, E.~Lunghi and S.~Shin,
``Hunting for Vectorlike Quarks,''
JHEP \textbf{04}, 019 (2019)
doi:10.1007/JHEP04(2019)019
[arXiv:1901.03709 [hep-ph]].


\bibitem{Buchkremer:2013bha} 
  M.~Buchkremer, G.~Cacciapaglia, A.~Deandrea and L.~Panizzi,
  ``Model Independent Framework for Searches of Top Partners,''
  Nucl.\ Phys.\ B {\bf 876}, 376 (2013)
  doi:10.1016/j.nuclphysb.2013.08.010
  [arXiv:1305.4172 [hep-ph]].

\bibitem{Degrande:2011ua}
C.~Degrande, C.~Duhr, B.~Fuks, D.~Grellscheid, O.~Mattelaer and T.~Reiter,
Comput. Phys. Commun. \textbf{183}, 1201-1214 (2012)
doi:10.1016/j.cpc.2012.01.022
[arXiv:1108.2040 [hep-ph]].

\bibitem{Alloul:2013bka} 
  A.~Alloul, N.~D.~Christensen, C.~Degrande, C.~Duhr and B.~Fuks,
  ``FeynRules  2.0 - A complete toolbox for tree-level phenomenology,''
  Comput.\ Phys.\ Commun.\  {\bf 185}, 2250 (2014)
  doi:10.1016/j.cpc.2014.04.012
  [arXiv:1310.1921 [hep-ph]].

\bibitem{Cacciapaglia:2010vn} 
  G.~Cacciapaglia, A.~Deandrea, D.~Harada and Y.~Okada,
  ``Bounds and Decays of New Heavy Vector-like Top Partners,''
  JHEP {\bf 1011}, 159 (2010)
  doi:10.1007/JHEP11(2010)159
  [arXiv:1007.2933 [hep-ph]].


\bibitem{Araque:2015cna}
J.~P.~Araque, N.~F.~Castro and J.~Santiago,
``Interpretation of Vector-like Quark Searches: Heavy Gluons in Composite Higgs Models,''
JHEP \textbf{11}, 120 (2015)
doi:10.1007/JHEP11(2015)120
[arXiv:1507.05628 [hep-ph]].
 
\bibitem{Dasgupta:2019yjm} 
  S.~Dasgupta, S.~K.~Rai and T.~S.~Ray,
  ``Impact of a colored vector resonance on the collider constraints for top-like top partner,''
  arXiv:1912.13022 [hep-ph].



\bibitem{Willenbrock:1986cr}
S.~S.~D.~Willenbrock and D.~A.~Dicus,
``Production of Heavy Quarks from W Gluon Fusion,''
Phys. Rev. D \textbf{34}, 155 (1986)
doi:10.1103/PhysRevD.34.155


\bibitem{Roe:2004na} 
  B.~P.~Roe, H.~J.~Yang, J.~Zhu, Y.~Liu, I.~Stancu and G.~McGregor,
  ``Boosted decision trees, an alternative to artificial neural networks,''
  Nucl.\ Instrum.\ Meth.\ A {\bf 543}, no. 2-3, 577 (2005)
  doi:10.1016/j.nima.2004.12.018
  [physics/0408124].

\bibitem{Romao:2019dvs}
M.~Romão Crispim, N.~F.~Castro, R.~Pedro and T.~Vale,
``Transferability of Deep Learning Models in Searches for New Physics at Colliders,''
Phys. Rev. D \textbf{101}, no.3, 035042 (2020)
doi:10.1103/PhysRevD.101.035042
[arXiv:1912.04220 [hep-ph]].


\bibitem{Roy:2020fqf} 
  A.~Roy, N.~Nikiforou, N.~Castro and T.~Andeen,
  ``Novel Interpretation Strategy for Searches of Singly Produced Vector-like Quarks at the LHC,''
  Phys.\ Rev.\ D {\bf 101}, 115027 (2020)
  doi:10.1103/PhysRevD.101.115027
  [arXiv:2003.00640 [hep-ph]].

\bibitem{Alwall:2011uj} 
  J.~Alwall, M.~Herquet, F.~Maltoni, O.~Mattelaer and T.~Stelzer,
  ``MadGraph 5 : Going Beyond,''
  JHEP {\bf 1106}, 128 (2011)
  doi:10.1007/JHEP06(2011)128
  [arXiv:1106.0522 [hep-ph]].

\bibitem{Cacciapaglia:2017iws} 
  G.~Cacciapaglia, G.~Ferretti, T.~Flacke and H.~Serodio,
  ``Revealing timid pseudo-scalars with taus at the LHC,''
  Eur.\ Phys.\ J.\ C {\bf 78}, no. 9, 724 (2018)
  doi:10.1140/epjc/s10052-018-6183-4
  [arXiv:1710.11142 [hep-ph]].

\bibitem{Wu:2014dba} 
  L.~Wu,
  ``Enhancing $thj$ Production from Top-Higgs FCNC Couplings,''
  JHEP {\bf 1502}, 061 (2015)
  doi:10.1007/JHEP02(2015)061
  [arXiv:1407.6113 [hep-ph]].


\bibitem{Cacciapaglia:2018qep} 
  G.~Cacciapaglia, A.~Carvalho, A.~Deandrea, T.~Flacke, B.~Fuks, D.~Majumder, L.~Panizzi and H.~S.~Shao,
  ``Next-to-leading-order predictions for single vector-like quark production at the LHC,''
  Phys.\ Lett.\ B {\bf 793}, 206 (2019)
  doi:10.1016/j.physletb.2019.04.056
  [arXiv:1811.05055 [hep-ph]].
  
  
\bibitem{Fuks:2016ftf} 
  B.~Fuks and H.~S.~Shao,
  ``QCD next-to-leading-order predictions matched to parton showers for vector-like quark models,''
  Eur.\ Phys.\ J.\ C {\bf 77}, no. 2, 135 (2017)
  doi:10.1140/epjc/s10052-017-4686-z
  [arXiv:1610.04622 [hep-ph]].



\bibitem{Sjostrand:2014zea} 
  T.~Sjöstrand {\it et al.},
  ``An Introduction to PYTHIA 8.2,''
  Comput.\ Phys.\ Commun.\  {\bf 191}, 159 (2015)
  doi:10.1016/j.cpc.2015.01.024
  [arXiv:1410.3012 [hep-ph]].

\bibitem{deFavereau:2013fsa} 
  J.~de Favereau {\it et al.} [DELPHES 3 Collaboration],
  ``DELPHES 3, A modular framework for fast simulation of a generic collider experiment,''
  JHEP {\bf 1402}, 057 (2014)
  doi:10.1007/JHEP02(2014)057
  [arXiv:1307.6346 [hep-ex]].

\bibitem{ATLAS:2016ukn}
 [ATLAS],
``Expected performance for an upgraded ATLAS detector at High-Luminosity LHC,''
ATL-PHYS-PUB-2016-026.





\bibitem{Atre:2011ae} 
  A.~Atre, G.~Azuelos, M.~Carena, T.~Han, E.~Ozcan, J.~Santiago and G.~Unel,
  ``Model-Independent Searches for New Quarks at the LHC,''
  JHEP {\bf 1108}, 080 (2011)
  doi:10.1007/JHEP08(2011)080
  [arXiv:1102.1987 [hep-ph]].

\bibitem{Atre:2013ap} 
  A.~Atre, M.~Chala and J.~Santiago,
  ``Searches for New Vector Like Quarks: Higgs Channels,''
  JHEP {\bf 1305}, 099 (2013)
  doi:10.1007/JHEP05(2013)099
  [arXiv:1302.0270 [hep-ph]].



\bibitem{Hocker:2007ht} 
  A.~Hocker {\it et al.},
  ``TMVA - Toolkit for Multivariate Data Analysis,''
  PoS ACAT {\bf }, 040 (2007)
  [physics/0703039 [PHYSICS]].




\end{thebibliography}
\end{document}